\documentclass[lettersize,journal]{IEEEtran}
\usepackage[frozencache,cachedir=.]{minted}
\usepackage{amsmath}
\usepackage{amssymb}
\usepackage{amsfonts}
\usepackage{algorithmic}
\usepackage{graphicx}
\usepackage{textcomp}
\usepackage{xcolor}
\usepackage{booktabs}
\usepackage[utf8]{inputenc}
\usepackage{tikz}
\usepackage{amsmath}
\usepackage{algorithm2e} 
\usepackage[listings]{tcolorbox}
\usepackage{xspace}
\tcbuselibrary{all}
\usepackage{pifont}
\usepackage{mathtools}
\usepackage{paralist}
\usepackage{tabularx}
\newtheorem{theorem}{Theorem}
\newtheorem{definition}{Definition}
\newtheorem{claim}{Claim}
\newtheorem{proof}{Proof}

\newcommand{\puppy}{\texttt{Puppy}\xspace}

\newcommand{\A}{\ensuremath{\mathcal{A}}\xspace}

\newcommand{\cache}{\ensuremath{\mathcal{C}}\xspace}

\newcommand{\Prov}{\ensuremath{\mathcal{P}}\xspace}
\newcommand{\Enclave}{\ensuremath{\texttt{TEE}}\xspace}
\newcommand{\FHE}{\ensuremath{\texttt{FHE}}\xspace}
\newcommand{\FE}{\ensuremath{\texttt{FE}}\xspace}
\newcommand{\MPC}{\ensuremath{\texttt{2PC}}\xspace}

\newcommand{\Ow}{\ensuremath{\mathcal{O}}\xspace}
\newcommand{\HD}{\ensuremath{\mathcal{H}}\xspace}

\newcommand{\secret}{\ensuremath{sec}\xspace}
\newcommand{\data}{asset\xspace}

\newcommand{\db}{\texttt{DB}\xspace}

\newcommand{\token}{token\xspace}
\newcommand{\tokens}{tokens\xspace}
\newcommand{\tokenH}{authorization \token}
\newcommand{\tokenP}{assignment \token}
\newcommand{\tokenPs}{assignment \tokens}

\newcommand{\np}{n_p}

\newcommand{\idtx}{id}

\newcommand{\FF}{\mathcal{F}}

\newcommand{\lrubase}{\ensuremath{\mathsf{LRU-Base}}\xspace}
\newcommand{\lrusim}{\ensuremath{\mathsf{LRU-Base-R}}\xspace}
\newcommand{\lruprop}{\ensuremath{\mathsf{LRU-Prop}}\xspace}

\newcommand{\abort}{\mathsf{abort}}

\newcommand{\PHash}{\ensuremath{\mathsf{PHash}}\xspace}
\newcommand{\hashparam}{\ensuremath{\mathsf{param}}\xspace}
\newcommand{\Sim}{\ensuremath{\mathsf{Sim}}\xspace}
\newcommand{\Get}{\ensuremath{\mathsf{GET}}\xspace}
\newcommand{\Put}{\ensuremath{\mathsf{PUT}}\xspace}

\newcommand{\Gen}{\mathsf{Gen}}
\newcommand{\Enc}{\ensuremath{\mathsf{Enc}}\xspace}
\newcommand{\Dec}{\ensuremath{\mathsf{Dec}}\xspace}

\newcommand{\SGen}{\ensuremath{\mathsf{Share}}\xspace}
\newcommand{\SReconst}{\ensuremath{\mathsf{Reconst}}\xspace}
\newcommand{\sharep}{\ensuremath{s_\Prov}\xspace}
\newcommand{\shareh}{\ensuremath{s_\HD}\xspace}

\newcommand{\pshare}{\ensuremath{tk_\Prov}\xspace}
\newcommand{\psharei}{\ensuremath{tk^i_\Prov}\xspace}
\newcommand{\hshare}{\ensuremath{tk_\HD}\xspace}
\newcommand{\hsharerandom}{\ensuremath{s'_\HD}\xspace}
\newcommand{\psharerandom}{\ensuremath{s'_\Prov}\xspace}

\newcommand{\pishare}{\ensuremath{s^{i}_{\Prov}}\xspace}

\newcommand{\IDGen}{\mathsf{IDGen}}

\newcommand{\Garble}{\ensuremath{\mathsf{Garble}}\xspace}
\newcommand{\Encode}{\ensuremath{\mathsf{Encode}}\xspace}
\newcommand{\GarbledEval}{\ensuremath{\mathsf{Eval}}\xspace}
\newcommand{\Decode}{\ensuremath{\mathsf{Decode}}\xspace}
\newcommand{\Evaluate}{\ensuremath{\mathsf{evaluate}}\xspace}
\newcommand{\circuit}{\ensuremath{f}\xspace}
\newcommand{\garbledCircuit}{\ensuremath{F}\xspace}
\newcommand{\encodingInfo}{\ensuremath{e}\xspace}
\newcommand{\decodingInfo}{\ensuremath{d}\xspace}
\newcommand{\garbledInput}{\ensuremath{X}\xspace}
\newcommand{\plainInput}{\ensuremath{x}\xspace}
\newcommand{\garbledOutput}{\ensuremath{Y}\xspace}
\newcommand{\plainOutput}{\ensuremath{y}\xspace}
\newcommand{\verifycircuit}{\ensuremath{\mathsf{Verify}}\xspace}

\newcommand{\SignKeyGen}{\ensuremath{\mathsf{KeyGen}}}
\newcommand{\Sign}{\ensuremath{\mathsf{Sign}}}
\newcommand{\SignVerify}{\ensuremath{\mathsf{Verify}}}

\newcommand{\WMKeyGen}{\ensuremath{\mathsf{SecretGen}}\xspace}
\newcommand{\WMGen}{\ensuremath{\mathsf{Insert}}\xspace}
\newcommand{\WMDet}{\ensuremath{\mathsf{Detect}}\xspace}

\newcommand{\uL}{u}

\newcommand{\sysone}{\textbf{SYS1}\xspace}
\newcommand{\systwo}{\textbf{SYS2}\xspace}
\newcommand{\systhree}{\textbf{SYS3}\xspace}
\newcommand{\secone}{\textbf{SEC1}\xspace}
\newcommand{\sectwo}{\textbf{SEC2}\xspace}
\newcommand{\secthree}{\textbf{SEC3}\xspace}

\newcommand{\Adv}{\ensuremath{\mathcal{A}}\xspace}
\newcommand{\B}{\ensuremath{\mathcal{B}}\xspace}

\newcommand{\Dist}{\ensuremath{\mathcal{S}}\xspace}
\newcommand{\Real}{{\mathsf{REAL}}}
\newcommand{\Ideal}{{\mathsf{IDEAL}}}

\newcommand{\Prot}[1]{\ensuremath{\Pi_{\scriptstyle\mathsf{#1}}}}
\newcommand{\Ppuppy}{\Prot{Puppy}\xspace}

\newcommand{\Func}[1]{{\FF_{\scriptstyle\mathsf{#1}}}}
\newcommand{\FSgx}{\Func{SGX}}

\newcommand{\Fpuppy}{\ensuremath{\Func{Puppy}}\xspace}

\newtcolorbox[auto counter,number format=\Roman ]{securitygame}[2][]{enhanced,colback=white,
fonttitle=\bfseries,coltitle=gray!25!black,
attach boxed title to top center={yshift=-3mm},,
boxed title style={colframe=gray!75!black,
colback=yellow!50!gray},
title=Security Game: #2,#1}
\newtcolorbox[auto counter,number format=\Roman ]{algorithmdes}[2][]{enhanced,colback=white,
fonttitle=\bfseries,coltitle=gray!25!black,
attach boxed title to top center={yshift=-3mm},,
boxed title style={colframe=gray!75!black,
colback=yellow!50!gray},
title=Algorithm: #2,#1}

\newtcolorbox[auto counter,number format=\Roman ]{functionality}[2][]{enhanced,colback=white,
fonttitle=\bfseries,coltitle=gray!25!black,
attach boxed title to top left=
{xshift=2mm,yshift=-3mm,yshifttext=-1mm},
boxed title style={colframe=gray!75!black,
colback=yellow!50!gray},
title=Ideal Functionality #2,#1}
\newtcolorbox[auto counter,number format=\Roman ]{puppyprotocolnonum}[2][]{enhanced,colback=white,
fonttitle=\bfseries,coltitle=gray!25!black,
attach boxed title to top left=
{xshift=2mm,yshift=-3mm,yshifttext=-1mm},
boxed title style={colframe=gray!75!black,
colback=yellow!50!gray},
title=#2,#1} %
\newtcolorbox[auto counter,number format=\Roman ]{puppyprotocol}[2][]{enhanced,colback=white,
fonttitle=\bfseries,coltitle=gray!25!black,
attach boxed title to top left=
{xshift=2mm,yshift=-3mm,yshifttext=-1mm},
boxed title style={colframe=gray!75!black,
colback=yellow!50!gray},
title=Protocol \thetcbcounter: #2,#1} %

\newcommand{\figlab}[1]{\label{fig:#1}}

\begin{document}

\title{\texttt{Puppy}: A Publicly Verifiable Watermarking Protocol}

\author{\IEEEauthorblockN{Devriş İşler\textsuperscript{1,2}, Seoyeon Hwang\textsuperscript{3}, Yoshimichi Nakatsuka\textsuperscript{4}\textsuperscript{\textsection}, Nikolaos Laoutaris\textsuperscript{1}, and Gene Tsudik\textsuperscript{3} }\\
\IEEEauthorblockA{\textit{\textsuperscript{1}IMDEA Networks Institute, \textsuperscript{2}UC3M, \textsuperscript{3}UC Irvine, \textsuperscript{4}ETH Zurich
}
}\\
\IEEEauthorblockA{{\small\{devris.isler, nikolaos.laoutaris\}@imdea.org, \{seoyh1, gene.tsudik\}@uci.edu, yoshimichi.nakatsuka@inf.ethz.ch}}}

\maketitle
\begingroup\renewcommand\thefootnote{\textsection}\footnotetext{Work done while the author was affiliated with UC Irvine.}
\endgroup

\begin{abstract}
In this paper, we propose \puppy, the first formally defined framework for converting \emph{any} symmetric 
watermarking into a publicly verifiable one.  \puppy allows anyone to verify a watermark any number of times  with the help of an {\bf untrusted} third party, without requiring owner presence during detection. 
We formally define and prove security of \puppy using the ideal/real-world simulation paradigm and construct two
practical and secure instances: (1) \puppy-\Enclave that uses Trusted Execution Environments (TEEs), and 
(2) \puppy-\MPC that relies on two-party computation (2PC) based on garbled circuits.
We then convert four current symmetric watermarking schemes into publicly verifiable ones and run 
extensive experiments using \puppy-\Enclave and \puppy-\MPC. Evaluation results show that,
while \puppy-\Enclave incurs some overhead, its total latency is on the order of milliseconds for three out 
of four watermarking schemes. Although the overhead of \puppy-\MPC  is higher (on the order of seconds),
it is viable for settings that lack a TEE or where strong trust assumptions about a TEE need to be avoided.
We further optimize the solution to increase its scalability and resilience to denial of service attacks via memoization.
\end{abstract}

\section{Introduction}\label{intro}
Data-driven technologies and machine learning are becoming ever more crucial to businesses and the entire society. A report from 
McKinsey~\cite{mckinsey16} predicts that the value of the data market will reach US$\$2.5$ trillion dollars by 2025. A wide variety of 
data types, including IoT real-time sensor data, financial, mobility, and health datasets, are used by the latest generation of machine 
learning-based services. To meet this demand, intermediaries that provide platforms to trade various data, so-called \textit{data 
marketplaces} (e.g., AWS, Refinitiv)~\cite{azcoitia2022survey}, play an increasingly 
important role in empowering the global data economy.
The major goal of data marketplaces (DMs) is to connect data buyers with data providers/sellers/owners and give them a reliable platform 
to trade their digital assets, e.g., machine learning models, datasets, and various other content. Initiatives, such as 
GAIA-X~\cite{gaiax} and IDSA~\cite{idsa} have been launched to define infrastructure and exchange standards for supporting this 
growing ecosystem.

The majority of data marketplaces and data owners surveyed in~\cite{azcoitia2022survey} explicitly quote a policy that the 
purchased goods cannot be resold or used outside of the pre-agreed jurisdiction of the buyer. 
However, after a \textit{digital asset} (hereafter referred to as \textit{asset}) is sold, it is hard to prevent the 
buyer from violating the agreed-upon policy. 
For instance, a malicious buyer can resell the \data, thus violating the terms and conditions under which the \data was purchased. 
Therefore, protecting ownership rights is crucial for the growth of the data economy.

\emph{Watermarking} is a well-known technique for protecting \data ownership rights by deterring unauthorized (re)distribution.
It supports various types of data, including audio, video~\cite{asikuzzaman2017overview}, 
images~\cite{anand2020watermarking,begum2020digital,liu2019optimized}, 
software~\cite{collberg2002watermarking,venkatesan2001graph}, 
databases~\cite{agrawal2003watermarking,agrawal2002watermarking,al2008robust,kamran2018comprehensive,shehab2007watermarking,sion2004rights},
other data types~\cite{ayday2019robust,jindss23,ji2022robust,panah2016properties,isler22}, and even (deep) neural 
networks~\cite{adi2018turning,chen2018deepmarks,darvish2019deepsigns,li2021towards,tekgul2021waffle,uchida2017embedding,wwwLiuLLWXWZ023,zhang2020model}.
A watermarking scheme consists of two algorithms: watermark \textit{insertion} and \textit{detection}. 
The owner uses the insertion algorithm to embed a (visible or invisible) watermark into an \data using a secret, 
called \textit{watermarking secret},
and it uses the same watermarking secret\footnote{
In this work, we only consider \textit{symmetric} watermarking, meaning that the \textit{watermarking secrets} used for insertion and detection are the same.
While there exist \textit{asymmetric} watermarking schemes~\cite{kumarecent2020,panah2016properties} where the secrets are different, we consider them out of scope.} during the detection algorithm to prove ownership (even if the \data is modified).
Due to the imperceptibility property of watermarking schemes, inserting watermarks does not affect the utility/functionality of the original \data.

In the context of data economy, a seller (owner) inserts a watermark into an \data before 
uploading it to a marketplace.  Later, if the seller suspects that a different \data on the marketplace
resembles one of its own,  it privately runs the detection algorithm using its watermark secret to check whether
someone is maliciously claiming the \data as theirs.
However, to prove ownership \emph{publicly} allowing DMs to deter infringement, a seller must reveal its watermarking secret.

\noindent \textbf{The problem:} Revealing the secret is problematic because anyone knowing the secret can remove the watermark, insert a new one, and stop a seller from proving its ownership of that \data in the future.
Therefore, it is commonly accepted~\cite{agrawal2002watermarking,isler22} that a seller can publicly prove ownership \emph{only once}. One way to address this problem is to involve a trusted third party (TTP), such as a certificate authority (CA), as follows: CA generates a certificate for the watermarking secret of the seller that can be distributed together with the watermarked \data (as proposed in~\cite{li2006publicly} for relational databases). However, this approach still discloses the secret, causing the watermarked \data to be still vulnerable to removal attacks.  
An alternative approach is to embed multiple watermarks and reveal one secret per detection~\cite{lach1999robust,adi2018turning}. This restricts the number of detections and does not prevent malicious entities from removing the watermarks after obtaining all corresponding secrets.
Also, this method may not be applicable to watermarking schemes that are vulnerable to ambiguity attacks~\cite{adelsbach2003ambiguity}. 

Some prior works~\cite{adelsbach2003watermark,adelsbach2001zero,adi2018turning,saha2011secure} propose letting the seller (as a rightful owner) prove that an \data is indeed watermarked without revealing any information about the watermark secret by  using zero-knowledge proofs (ZKPs)~\cite{goldreich1994definitions,goldwasser1989ZK,menezes2018handbook}. 
ZKPs allow the seller to prove ownership any number of times. 
However, this requires the seller to be online for watermark detection because the detection algorithm requires the watermark secret in order to generate a zero-knowledge proof. 
If the detection is conducted frequently, the seller can be overwhelmed due to the high computational cost of ZKPs on the detection phase.
This can lead to poor performance and scalability issues in DMs with numerous sellers and buyers. 

In summary, the aforementioned solutions for publicly verifiable watermarking exhibit at least one of the following shortcomings: 
(1) limited number of (public) watermark detections, (2) restricted applicability to current watermarking techniques, 
(3) requires seller involvement during watermark detection, and (4) high computational cost. 
This prompts us to consider the following question:
\begin{quote}
\centering

\emph{Can any symmetric watermarking scheme be converted into a publicly verifiable one without 
incurring any of the above shortcomings?}

\end{quote}

\noindent \textbf{Our approach:} This paper tackles the above challenge by proposing \puppy, the first 
\emph{formally defined} framework that converts \emph{any} symmetric watermarking scheme into a \emph{publicly verifiable} one.
\puppy consists of two phases: \emph{ownership generation} and \emph{verification}.
During the former, the seller runs the watermark insertion algorithm 
and generates two \tokens using the watermarking secret: (1) \tokenH given to the buyer (called \textit{Holder}) 
together with the purchased \data, and (2) \tokenP given to an \textit{untrusted} third party, called \textit{Prover}.
In verification phase, \textit{Holder} and \textit{Prover} engage in a secure computation protocol that: 
(1) runs the detection algorithm on the \data using their \tokens, and (2) returns the detection result to \textit{Holder}.
This way, the \textit{Holder} can perform verification (i.e., watermark detection) any number of times. 

\puppy is \emph{agnostic} with respect to the underlying secure computation framework, and can be implemented using various 
primitives, such as Fully Homomorphic Encryption (FHE), Secure Two-Party Computation (2PC), Functional Encryption (FE), or
Trusted Execution Environments (TEEs).  
We investigate multiple instances of \puppy framework based on these primitives and compare them in several aspects. 
We then demonstrate two concrete instances that meet our requirements: one that relies on TEEs, called \puppy-\Enclave, and the other that uses a 2PC protocol, called \puppy-\MPC. 
\puppy operators can choose between a high-performing TEE instance and a provably-secure 2PC instance, according to their needs. 
We emphasize that selecting underlying primitives is not trivial, as discussed in Section \ref{subsec:secure_computation_instances}, since it requires careful consideration of many alternatives in order to satisfy security and performance goals.

To demonstrate \puppy's \emph{versatility} and \emph{practicality}, we convert four symmetric watermarking techniques into publicly verifiable ones using \puppy-\Enclave: (1) dataset~\cite{isler22}, (2) numerical relational database~\cite{shehab2007watermarking}, 
(3) image~\cite{lou2007copyright}, and (4) Deep Neural Network (DNN)~\cite{adi2018turning}. 
\puppy-\Enclave is implemented using the well-known Intel SGX~\cite{mckeen2013innovative}, and \puppy-\MPC is built upon
the \texttt{MP-SPDZ} framework~\cite{mp-spdz}.

Beyond providing the desired functionality, \emph{performance} is important for \puppy, since real-world marketplaces need to support 
high verification rates. Furthermore, since Holder may have limited computational power, \puppy avoids computationally heavy tasks on Holder-side.
Our evaluation results show that, for three out of four watermarking techniques, \puppy-\Enclave introduces minor latency overhead during the verification and the overall latency is in the order of milliseconds.
Only the DNN instance incurs overhead that results in 1-2 orders of magnitude slowdown.
We also converted the dataset watermarking scheme using \puppy-\MPC and showed that, despite performance 
challenges, \puppy-\MPC can convert private watermarking techniques into public ones.

Finally, we further \emph{optimize} \puppy's performance by introducing memoization.
It enables provers to cache previously computed verification results without degrading security. 
This allows provers (and holders) to completely skip the verification procedure for similar assets 
determined via perceptual (e.g., locality-sensitive) hashing.
The proposed optimization technique is based on a new \emph{proportional caching mechanism}. 

The contributions of this work are as follows: \\
\noindent $\bullet$ \puppy framework that: (1) converts \emph{any} symmetric watermarking scheme into a publicly verifiable one, (2) does not require owner involvement during watermark detection, (3) allows unlimited number of watermark detections, and (4) is efficient.\\
\noindent $\bullet$ Two proof-of-concept implementations and their comprehensive security 
analysis: (1) \puppy-\Enclave using Intel SGX; and (2) \puppy-\MPC using Yao's Garbled Circuits.\\
\noindent $\bullet$ Extensive evaluation of \puppy-\Enclave performance using four types of symmetric watermarking schemes and experimental analysis on the optimized version. 

\section{Background}\label{prelim}
\subsection{Symbols and Notation}
Let $\lambda \in \mathbb{N}$ be a security parameter. A probabilistic polynomial time (PPT) algorithm is a probabilistic algorithm taking $1^\lambda$ as 
input that has running time bounded by a polynomial in $\lambda$. A positive function $negl: \mathbb{N} \rightarrow \mathbb{R}$ is called 
\textit{negligible}, if for every positive polynomial $poly(\cdot)$, there exists a constant $c > 0$ such that for all $x > c$, we have  $negl(x) < 
1/poly(x)$. $||$ denotes concatenation.  
We denote a function executed by a single party as $\texttt{output}\leftarrow \texttt{Function(input)}$.
All other notation is shown in Table~\ref{tab:notation}.

\subsection{Watermarking Primer}\label{wmprimer}
A watermarking scheme covertly embeds invisible data (a watermark) into a given \data (or signal), using a watermarking secret. It includes three 
PPT algorithms:\\
\noindent $\bullet$ $\WMKeyGen(1^\lambda)$ generates a secret \footnote{Here, we refer any secret information needed for generating and verifying a watermarked \data. e.g., a watermarking key \cite{agrawal2002watermarking,agrawal2003watermarking}, a trigger set \cite{adi2018turning}, the pixel positions of marks that are inserted in an image data.} $\secret$; \\ 
\noindent $\bullet$ $\WMGen(D_o,\secret)$ inserts a watermark on given \data $D_o$ and generates a watermarked \data $D_w$ using $\secret$. Note that $D_w$ and $D_o$ are perceptually similar; and\\
\noindent $\bullet$ $\WMDet(D_w,\secret)$\footnote{\WMDet algorithm generally depends on a threshold value for acceptance.} either \textit{accepts} or \textit{rejects} the claim that the given $D_w$ is a valid watermarked \data using $\secret$. 

We assume that a symmetric watermarking scheme is secure and robust, according to the standard definitions proposed for the given \data type.
A watermarking scheme is assumed to be secure against the guessing attack (where an attacker tries to expose the watermarking secret) and robust against intentional/unintentional alterations/modifications (i.e., a watermark should be still detectable even under attacks such as~\cite{agrawal2003watermarking,agrawal2002watermarking,barak2001possibility,bas2009two,cohen2018watermarking,ji2021curse,quiring2018forgotten,shafieinejad2021robustness}). 
However, depending on the type of \data,
additional security requirements may be needed, such as unremovability, unforgeability, and ambiguity~\cite{adelsbach2003ambiguity}.\footnote{Watermarking is assumed to be secure and robust against the attacks defined in its domain. For instance, most digital watermarking schemes require to be secure against ambiguity attacks while relational database watermarking methods do not impose such requirements due to non-interference property.} 
We formally define common properties of watermarking which we are inspired from \cite{adi2018turning,barak2001possibility} in Appendix \ref{wmfurther}. 

\subsection{Secret Sharing} 
Secret sharing (SS) is a cryptographic primitive that allows a dealer to distribute a secret $s$ among multiple parties such that any authorized set of parties can reconstruct $s$, whereas an adversary corrupting the other sets of parties learns nothing about $s$. 
The pieces of $s$ each party holds are called ``shares'' and the collection of the shares is called the ``sharing'' of $s$. 
Typically, a SS scheme consists of two PPT algorithms:\\
\noindent $\bullet$ \SGen$(s, P)$ generates a sharing of $s$, given secret $s$, and distributes them to the parties in $P$ according to the authorization rules. $s_{P_i}$ denotes the share(s) that the party $P_i \in P$ holds. \\
\noindent $\bullet$ \SReconst$(\{s_i\}_{i \in B})$ reconstructs the secret using the input shares held by parties in $B$. It outputs the secret $s$ if $B$ is an authorized set, or $\perp$ otherwise.

The authorized set varies depending on the scheme and the adversary model. For example, in Shamir's SS scheme~\cite{shamir1979share}, any $t$ or more parties can reconstruct $s$ using their shares for a threshold value $t$, whereas the additive secret sharing scheme~\cite{Maurer02additiveSS} requires all shares to reconstruct $s$.
    
\subsection{Trusted Execution Environments}\label{teeprel}
A Trusted Execution Environment (TEE) is a security primitive that isolates code execution and data from untrusted code (see Appendix \ref{sgxfuncdef} for its formal definition).
A typical TEE provides the following features:\\
\noindent $\bullet$ \textbf{Isolated Execution.} TEEs provide an execution environment for the security-sensitive code and data, isolated from all other software on the platform, including OS, hypervisor, and BIOS. The data inside the TEE can be accessible only by the code running inside the TEE.\\
\noindent $\bullet$ \textbf{Controlled Invocation.} The code inside the TEE can only begin and end from pre-defined entry/exit points (e.g., call gates) which are enforced by the TEE.\\  
\noindent $\bullet$ \textbf{Remote Attestation.} A TEE generates a cryptographic proof to a remote party that assures that the TEE is genuine and the code running within the TEE is the one expected. 
The remote party can use this assertion to decide whether to trust the TEE or not and then establish a secure communication channel with it.

There are several instances of TEE, including Intel Software Guard Extensions (SGX)~\cite{mckeen2013innovative}, ARM TrustZone~\cite{ARMtrustzone}, and AMD Secure Encrypted Virtualization (SEV)~\cite{kaplan2016amd}.
\subsection{Secure 2-Party Computation (2PC)}\label{2pcgc}
A secure 2-party computation \cite{evans2018pragmatic} allows two parties that do not trust each other to compute a function 
$f(x_1,x_2)$ jointly where $x_1$ and $x_2$ are private inputs of the parties. 
Parties learn nothing more than the function result.
In our work, we focus on 2PC utilized by Yao's garbled circuit (GC) \cite{yao86} (see Appendix \ref{garbledef} for its formal definition) and Oblivious 
Transfer (OT)~\cite{yadav2021survey} (see Appendix \ref{otdef} for its description).
In a garbled circuit-based 2PC protocol, the function \circuit is represented as a (boolean) circuit with $2$-input gates (e.g., XOR, AND, etc.). 
One of the parties, called \emph{garbler}, garbles the circuit by assigning two random keys to each wire in the circuit. 
For each gate, a garbled table is computed that allows decryption of the gate’s output key given its two input keys. 
The garbler then sends the encrypted circuit (\garbledCircuit) along with its corresponding input keys to another party, the \emph{evaluator}. 
To evaluate \circuit, the evaluator needs to acquire its input keys without revealing its actual input data to the garbler. 
To do so, the evaluator obtains its input keys obliviously through a $1$-out-of-$2$ OT protocol and uses them to evaluate the garbled circuit gate by gate. 
Finally, the evaluator decrypts the encrypted output (e.g., watermark verification result $res$) using the decoding information (i.e., the mapping keys).

\begin{table}[htp]
\centering
\caption{Summary of Notation.}
\begin{tabular}{|c|c|}
\hline
$D_o$         &   Original \data \\ \hline 
$D_w$ &  Watermarked \data  \\ \hline
$\idtx$&  Identifier of $D_w$  \\ \hline
\secret & Watermarking secret \\ \hline
\Ow         &  Owner of $D_o$  \\ \hline
\HD         &   Holder of $D_w$   \\ \hline
\Prov         &  Prover       \\ \hline
$\hshare$ & Authorization token to \HD \\ \hline
$\pshare$ & Assignment token to \Prov \\ \hline
\end{tabular}
\label{tab:notation}
\end{table}

\section{System \& Threat Model} \label{overview}
\subsection{System Model} \label{subsec:system_design}
An \data\ refers to \emph{any} type of data that can be watermarked (e.g., digital objects, datasets, ML models). 
Our system model involves three kinds of entities:\\
\noindent $\bullet$ \textit{\bf Owners:} who sell some \data $D_o$ to \textit{Holders} after watermarking the \data ($D_w$), using a watermark secret, $\secret$;\\
\noindent $\bullet$ \textit{\bf Holders:} who obtain/buy watermarked \data\ $D_w$ from an \textit{Owner} and verify its validity (i.e., ownership of $D_w$); and \\
\noindent $\bullet$ \textit{\bf Provers:} who help \textit{Holders} to verify the validity of $D_w$. 

For the sake of clarity, we assume one party for each role, denoted as \textit{Owner} (\Ow), \textit{Holder} (\HD), and \textit{Prover} 
(\Prov), unless explicitly specified otherwise. 
Note that \puppy can easily be modified to support multiple \textit{Holders} (with the same watermarked \data or a different watermarked 
\data for each \textit{Holder}) or multiple \textit{Provers}, which we discuss in Section~\ref{further}. 
We assume that \Ow and \HD are entities with limited computing power (e.g., consumer-grade laptop), 
whereas \Prov is assumed to have more powerful computation resources (e.g., cloud server) to provide services to \HD.
Additionally, \Ow is required to be online only during \data sharing/selling to \HD and goes offline after that.

\subsection{Adversary Model}\label{advmodel}
We consider a static adversary \A that can corrupt either \HD or \Prov, though not both at the same time. 
This means that \A chooses a party to corrupt before the protocol begins, and cannot change its choice once the protocol starts. 
The corrupted party can arbitrarily deviate from the protocol and \A can observe all of its state information, 
including secrets. 
We assume \Ow remains honest.
The scenario with with multiple \HD-s and multiple \Prov-s is discussed in Section~\ref{further} and the case with 
corrupted \Ow-s -- in Section \ref{futurework}. 
The goal of $\A$ corrupting \HD is to obtain \secret used in $D_w$ and perhaps use that secret to remove the watermark. 
On the other hand, $\A$'s goal in corrupting \Prov may be to generate false verification results, learn the identity of the owner of $
D_w$, and/or obtain \secret and $D_w$.

Note that \puppy does not tackle the problem/issue of revealing statistical information about verification requests.
We consider it to be a separate problem.

We assume \Ow uses secure authenticated channels to communicate with \HD and \Prov, while \HD and \Prov use an 
anonymous communication channel~\cite{ando2022poly} (e.g., Tor~\cite{dingledine2004tor}) for anonymity.

All underlying cryptographic primitives used in \puppy instances are assumed to be secure.
Online attacks, such as denial-of-service, are not the main focus of our work, although they can be mitigated by well-known techniques, such as CAPTCHAs~\cite{ahn2003captcha,gossweiler2009s,nakatsuka2021cacti,singh2014survey} or via our memoization approach discussed in Section \ref{memoization}.

\subsection{System \& Security Requirements}\label{requirement}
The system and threat model described above yield the following system (\textbf{SYS}) and security (\textbf{SEC}) requirements: \\
\noindent $\bullet$ \textbf{\sysone. Owner-free Verification:} \HD can verify $D_w$ without interacting with \Ow. \\
\noindent $\bullet$ \textbf{\systwo. Unlimited Verification:} \HD can verify $D_w$ any number of times. \\
\noindent $\bullet$ \textbf{\systhree. Holder Friendliness:} \HD can run the protocol without requiring high computational power. \\
\noindent $\bullet$ \textbf{\secone. Secret Protection:} No one, except \Ow, should learn $sec$. \\  
\noindent $\bullet$ \textbf{\sectwo. Watermarked Asset Protection:} No one, except \HD, should learn anything about $D_w$. \\
\noindent $\bullet$ \textbf{\secthree. Privacy:} All transaction history between \Ow and \HD must be kept private from any third parties.  

\section{\puppy Design} \label{design}
This section begins with some na\"ive approaches to public verification, identifies key challenges, and then describes how \puppy overcomes them.
To demonstrate its flexibility, we provide several \puppy\ instances that use very different cryptographic techniques and components.
Next, we consider two specific instances: \puppy-\Enclave and \puppy-\MPC.

\subsection{Na\"ive Approach}\label{subsec:strawman} 
One straightforward approach to convert a symmetric watermarking scheme to a publicly verifiable one is to outsource watermark 
detection to a third party (TP), such as a cloud server. This way, \Ow generates $D_w$ using \secret, and sends \secret to TP 
which interacts with \HD and returns the verification result. 
However, this violates \secone and \sectwo, since TP acquires both \secret and $D_w$, allowing it to remove the watermark 
from $D_w$. TP also tracks all communication between \Ow and \HD, which violates \secthree. 

Another approach is to embed multiple secrets (as in~\cite{adi2018turning,lach1999robust}) on one \data, revealing only one at a time, whenever a verification is requested. 
This can be achieved by the non-interference property, i.e., if watermarks do not interfere with each other.
However, this approach violates \secone, \sysone, and \systwo: \Ow's secret is revealed to \HD; \Ow must be online in the 
verification phase; and a limited number of verification operations can be performed.

\subsection{Design Challenges}\label{subsec:challenges}
Intuitive approaches described above highlight several challenges in converting a symmetric watermarking scheme into a publicly verifiable one:

\noindent\textbf{Offline \Ow.}
We cannot assume \Ow being always online.
Suppose that \HD receives a watermarked \data from \Ow and later realizes that it is very similar to some previously received \data.
If the watermark detection algorithm requires \Ow's participation, \HD needs to wait for 
\Ow to be online. This hinders processing of \data and introduces additional communication delay between \Ow and \HD.

\noindent\textbf{Handling the secret.}
In a symmetric scheme, the same secret is used for both inserting and detecting a watermark.
It must be protected from all other entities meaning that its handling/storage is a challenge.

\noindent\textbf{Computation using the secret.}
Even if the secret is otherwise safe, it is still 
needed for watermark detection.
Securing it during watermark detection is another challenge.

\noindent\textbf{Size of the secret.}
The sizes of the secret in various watermarking schemes are not always small.
For instance, the symmetric secret in one of the DNN watermarking schemes is on the order of 
tens of MBs; in other schemes, it is around several KBs (See Section~\ref{overhead} for more information).
If the secret size is large, it can be computationally intensive for the parties involved while handling the secret and running watermark detection using the secret. 

\subsection{Realizing the Design}\label{subsec:realizing}
This section describes how \puppy addresses the aforementioned design challenges.

\textbf{Offline \Ow.}
\puppy does not require \Ow to be online during watermark detection.
However, \Ow must provide potentially untrusted entities with its watermarking secret.  
While doing so, \puppy shall not inherit the challenges of FHE/FE-based approaches.
To mitigate this problem, \Ow grants \HD a right to compute watermark detection whilst \Ow is still the only party having the full knowledge of the watermarking secret. 
Below, we discuss how \puppy achieves it.

\textbf{Handling the secret.}
We introduce a new \emph{untrusted} entity -- \emph{Prover} (\Prov) -- that assists \HD during watermark detection.
\Ow encrypts the watermarking secret with a symmetric key (more on this later) and derives two tokens (\emph{\tokenH} and \emph{\tokenP}) from the key such that it is computationally infeasible to learn any information about the secret from either or both token(s).
By giving \tokenH along with the watermarked \data to \HD, \Ow \textit{authorizes} \HD to verify ownership of watermarked \data.
On the other hand, \Prov is \textit{assigned} as \textit{Prover} to assist \HD by receiving \tokenP.
This way, the watermarking secret is not held/revealed by any entity (other than \Ow) at any point of time, except during watermark detection, described below.

\textbf{Computation using the secret.}
We use \emph{secure computation} for  \HD to verify the presence of a watermark without knowing the watermarking secret.
The main idea is to enforce \HD and \Prov to privately reconstruct the watermarking secret using their respective tokens during the watermark detection algorithm. 
Candidate secure computation mechanisms are discussed in the next section.

\textbf{Secret Size.}
\Ow encrypts the watermarking secret with a fresh symmetric key and derives two tokens from the symmetric key.
With this procedure, the size of tokens stored by \HD and \Prov is fixed forming token sizes independent of secret size. 
This enables \HD and \Prov to exchange a small amount of information yielding low communication and computation overhead while executing a \puppy instance. 

\begin{figure}[htp]
\centering
\includegraphics[width=0.5\textwidth]{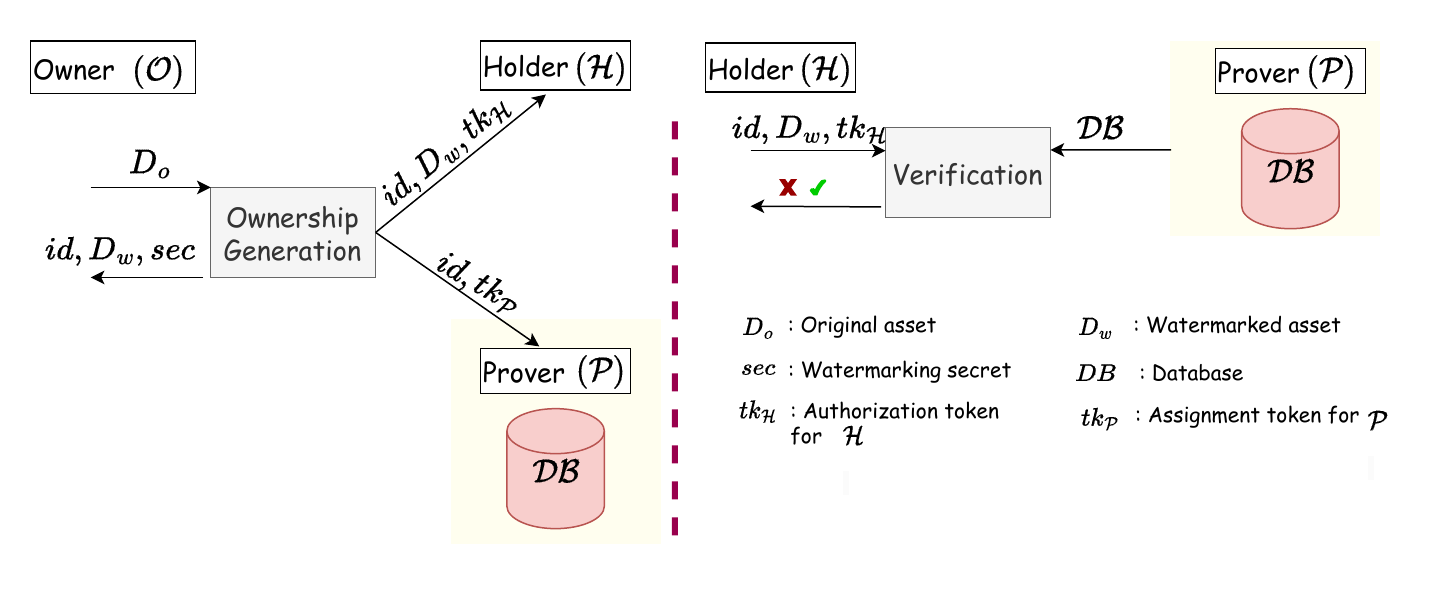}
\caption{\protect \puppy Overview.}\label{fig:puppy}
\end{figure}

\subsection{\protect \puppy Framework} \label{subsec:puppy_framework}
We now overview the \puppy framework, which consists of two phases: \textit{ownership generation} and \textit{verification}, 
as shown in Figure~\ref{fig:puppy}. In the former (denoted by \texttt{Generate} and formally defined in Appendix \ref{formaldef}),
\Ow generates a watermarked \data, $D_w$, and a token-pair: \emph{\tokenH} and \emph{\tokenP}. 
Specifically, \Ow inserts a watermark into the original \data, $D_o$, using \secret, and obtains the watermarked \data, 
$D_w$. For public verifiability, \Ow generates (1) \tokenH ($tk_\HD$) that authorizes \HD to verify ownership of $D_w$ without \Ow's help, and  (2) \tokenP ($tk_\Prov$) that assigns \Prov to be a \textit{Prover} which helps \HD to later verify ownership of $D_w$ 
Next, \Ow distributes $tk_\HD$ and $tk_\Prov$ to each party with an identifier $\idtx$ for \data.
We assume that \Ow and \HD agreed to buy/sell \data beforehand, potentially for monetary payment~\cite{azcoitia2022survey}, or for some other non-monetary incentive.  

In the verification phase (denoted by \texttt{Verify} and formally defined in Appendix \ref{formaldef}), \Prov and \HD engage in a secure computation instance, whereby
\Prov helps \HD verify ownership of $D_w$ without learning anything about \secret or $D_w$.
\Prov manages a database \db for collecting all identifiers and \tokenPs (received from \Ow-s) and uses it in the verification phase.
In the end, \HD learns whether ownership of $D_w$ is valid (denoted by `1') or not (denoted by `0'), while \Prov learns nothing.

\subsection{Deriving \puppy Instances with Secure Computation Techniques}\label{subsec:secure_computation_instances}
There are several well-known secure computation techniques.
We now discuss whether and how they can be integrated into the \puppy framework. 
Results are summarized in Table~\ref{tab:compinstances}.
Note that we only consider one \HD and one \Prov for simplicity.

\noindent\textbf{Fully Homomorphic Encryption (FHE)~\cite{chillotti2020tfhe,gentry2009fully}}: One party (\HD or \Prov) outsources its \token, encrypted under its own public key, to the other party, and the latter
executes the watermark detection algorithm using its \token and the received token, both encrypted using an FHE under the first
party's public key. For example, \HD sends a message to \Prov, asking for verification by sending an identifier.
\Prov encrypts corresponding \tokenP under its public key and sends it to \HD.
\HD runs the detection algorithm on encrypted \tokenP using its \tokenH and watermarked \data, both encrypted with \Prov's public key.
Then, \HD sends the encrypted result to \Prov which can decrypt it.
As mentioned earlier, we assume an anonymous communication channel is used 
between \HD and \Prov for \secthree to hide the identity of \HD from \Prov. 

Despite impressive recent advances in FHE, its computation cost is still quite high for arbitrary 
(especially complicated) functions~\cite{viand2021sok}.
This is especially problematic for \puppy as it requires \HD to do most of the heavy-lifting despite having lower computational power compared to \Prov.
Furthermore, advanced technologies are needed to handle malicious \Prov, such as zero-knowledge proofs (ZKPs) 
and message authentication codes (MACs), which incur further computation and communication costs. 

Also, moving the computation to \Prov is not trivial.
\Prov needs \HD's public key to encrypt its \token. However, this key allows \Prov to tie a certain \tokenP to 
\HD, violating \secthree.
Therefore, \Prov needs to somehow retrieve an \tokenP without learning the identity of \HD.

\noindent \textbf{Functional Encryption (FE)~\cite{boneh2011functional}}:
Using FE, \Prov generates necessary keys (e.g., functional secret key) and uses them to encrypt its \tokenP. 
Then it sends the ciphertext -- along with the functional secret key and its public key -- to \HD.
\HD runs the FE decryption algorithm using the functional secret key, encrypted \tokenP, and its \tokenH, thus
allowing it to obtain the result of the detection algorithm.

Similar to FHE, the FE approach requires \HD to run the FE decryption algorithm to meet \secthree,
as \secthree would be violated if \Prov runs the FE decryption algorithm.
However, this induces prohibitive computational costs for 
\HD, which violates \systhree.

\noindent \textbf{Secure Multi-Party Computation (SMPC)~\cite{evans2018pragmatic}}:
We focus on the 2-party case, i.e., 2PC. 
In a one-\HD and one-\Prov setting, \HD and \Prov engage in a 2PC instance using their \tokens.
In the end, \HD receives the result of the detection algorithm, while \Prov receives nothing.
Depending on the specific circuit, 2PC may require online communication between \HD and \Prov, which incurs certain 
communication overhead for \HD.

\noindent \textbf{Trusted Execution Environments (TEEs), e.g., Intel SGX~\cite{mckeen2013innovative}}:
With a TEE-based approach, \HD sends its \tokenH to the TEE on \Prov after performing remote attestation.
\Prov sends its corresponding \tokenP to TEE.
The TEE computes the watermark detection algorithm using \tokenP and \tokenH and returns the result to \HD.
\Prov learns nothing.

\noindent \textbf{Summary}: While both FHE-based and FE-based \puppy
instances satisfy all of our security requirements, they do not
satisfy \systhree, and are thus unsuitable for an \HD
with relatively low computational power. Hence, for the rest of the paper, we focus on instantiating \puppy via TEE and 2PC approaches.

\begin{table}[htp]
    \centering
    \caption{Comparison of \puppy instances based on each primitive: \Enclave, \MPC, \FHE, and \FE. 
    \HD Comp. indicates the computational overhead on \HD, while \Prov Comp. indicates the computational overhead on \Prov. \textit{Comm. Overhead} indicates the communication overhead between \HD and \Prov.}
    \begin{tabular}{l|c|c|c}
    \hline
        \textit{Primitive} & \textit{\HD Comp.} & \textit{\Prov Comp.} & \textit{Comm. Overhead} \\ \hline \hline
        \Enclave & Low       & Low         & Low                \\ \hline
        \MPC     & Moderate  & Moderate    & Moderate/High          \\ \hline
        \FHE     & High      & Moderate    & Low/Moderate       \\ \hline
        \FE      & High      & Moderate    & Low/Moderate       \\ \hline
    \end{tabular}
    \label{tab:compinstances}
\end{table}

\subsection{TEE-Aided Construction (\puppy-\Enclave)}  \label{subsec:puppytee}
This section motivates and describes a TEE-based instance, denoted by \puppy-\Enclave.
\subsubsection{Use of \Enclave{s}} 
\Enclave{s} have three benefits as described below.

\textbf{Holder Friendliness.} 
Holders might not have sufficiently high computational power. Therefore, offloading the computationally-heavy watermark detection algorithm to Prover is desirable.

\textbf{Versatility.}
Since \puppy's main goal is to convert a symmetric watermarking scheme into a publicly verifiable one, supporting different watermarking techniques should be (ideally) effortless.
As discussed in Section~\ref{performance}, a TEE facilitates this since it can run arbitrary code  as long it is compiled for the correct CPU architecture.

\textbf{Performance.}
Table~\ref{tab:compinstances} provides an overview of the computation/communication overhead of the four \puppy instances.
This shows that, compared to other alternative solutions relying on cryptographic primitives, TEEs do not introduce much overhead.

\subsubsection{Ownership Generation Phase}
During this phase, \Ow first runs \WMKeyGen algorithm to generate \secret according to the security parameter $\lambda$. 
Then, \Ow generates $D_w$ using \secret via \WMGen.\footnote{
For the watermarking schemes that output \secret as a result of watermarking generation algorithm, the protocol can be easily modified to have $\{D_w,\secret\} \leftarrow \WMGen(D_o,1^\lambda)$ instead of steps (1a) and (1c).}
For this construction, we use Symmetric Encryption (SE) scheme (see Appendix \ref{sec:primitives} for the SE's full description) and Secret Sharing to generate the two tokens, where \tokenH is denoted as $\hshare$ and \tokenP as $\pshare$. 
As stated previously, due to the size of \secret, \Ow encrypts its watermarking secret under a secret key $k$ generated from a SE $\Gen$ scheme with respect to the security parameter $\lambda$.
\Ow encrypts the \secret using the $\Enc$ scheme under the key $k$ resulting $c_{sec}$.
Then \Ow secret shares the encryption key $k$ creating two shares: \shareh and \sharep.
This reduces the size of transmitting shares from the size of \secret to the size of $k$. 
Then, $\hshare$ and $\pshare$ are assigned as \shareh and $\langle \sharep, c_{sec}\rangle$, respectively.\footnote{ Note that which shares will be assigned as which tokens depends on the authorization rule defined by the secret sharing. For instance, for Shamir's secret sharing, it does not matter as in our simple construction.}

\Ow generates $\idtx$ for $D_w$ as well as these two shares, and then sends $\idtx$, $D_w$, and $\hshare$ to \HD, and $\idtx$ and $\pshare$ to \Prov.
\Prov stores $(\idtx,\pshare)$ in a database, \db, located outside of the \Enclave.
Protocol~\ref{ownershipTEE} depicts the ownership generation phase.

Note that \Ow and \HD may agree on a license agreement/policy which defines terms and conditions such as usage constraints, penalties in case of violations, and the rights that \Ow and \HD have.
We assume that this takes place when \HD requests $D_o$.

\begin{puppyprotocol}[label=ownershipTEE]{\puppy-\Enclave Ownership Generation}
On input of a \data $D_o$ from \Ow, the ownership of \Ow for $D_o$ is generated as follows:
\begin{enumerate}
  \item \Ow computes:
  \begin{compactenum}
    \item  $\secret \leftarrow \WMKeyGen(1^\lambda)$
    \item  $k \leftarrow \Gen(1^\lambda)$
    \item  $D_w \leftarrow \WMGen(D_o,\secret)$
    \item  $c_{sec} \leftarrow \Enc(k,\secret)$
    \item  $( \shareh,\sharep) \leftarrow \SGen(k,\{\HD,\Prov\})$
    \item  $\idtx \leftarrow \IDGen(D_w)^{*}$ 
  \end{compactenum}
  \item \Ow $\rightarrow$  \HD : $\langle \idtx,\hshare:=\shareh,D_w \rangle$
  \item \Ow $\rightarrow$  \Prov : $\langle \idtx,\pshare:=\{ \sharep,c_{sec}\} \rangle$
  \item \Prov: \db.\textbf{Store}($\idtx,\pshare$)
\end{enumerate}
\end{puppyprotocol}

\noindent \textit{(*) Note. Generation of $\idtx$ by $\IDGen(.)$}: 
$\idtx$ is an identifier for \Prov, which allows it to later search the share needed for the verification request of \HD. 
Therefore, $\idtx$ must be unique for each watermark. 
For example, it can be defined as $\IDGen(D_w) = H( \Ow || M_{D_w} || date)$, where $H$ is a cryptographic hash function and $M_{D_w}$ is the metadata of $D_w$.
\Ow could also use a pseudo-random function to generate a random $\idtx$, using $D_w$ as a seed.  

\subsubsection{Verification Phase}
During the verification phase (as shown by Protocol~\ref{verificationTEE}), \HD verifies the watermark of \Ow using $D_w$ and $\hshare$ with $\idtx$.
First, \HD performs a remote attestation ($\mathsf{RA}$) on \Prov.\Enclave over the anonymous communication channel (e.g., Tor).
It aborts if the attestation fails. 
Otherwise, \HD establishes a secure channel with \Prov.\Enclave, and from this point on, all messages between \HD and \Prov.\Enclave are sent through this secure channel.
\HD sends $\langle \idtx$, $D_w$, and $\hshare \rangle$ to \Prov.\Enclave.
Once \Prov.\Enclave receives the request, it obtains \pshare consisting of $\sharep$ and $c_{sec}$ from the \db using $\idtx$. 
It then reconstructs the decryption key $k$ from the shares (via \SReconst) where shares (\shareh and \sharep) are parts of the tokens (\hshare and \pshare). 
Later on, it decrypts $c_{sec} \in \pshare$ resulting in \secret. 
After decryption, it runs \WMDet with $\secret$ and $D_w$, and obtains the result $res$.
\Enclave replies with $\langle res \rangle$ to \HD. 
\Enclave then erases all data including the watermarked \data, secret shares, decryption key, and watermarking secret(s). 
Finally, \HD returns \texttt{Valid} if $res$ is `1' or \texttt{Invalid} otherwise. 

\begin{puppyprotocol}[label=verificationTEE]{\puppy-\Enclave Verification}
On input of $\idtx$, $D_w$, and $\hshare$ from \HD, and \db from \Prov, the verification phase is as follows: 
\begin{enumerate}
  \item \HD $\xLeftrightarrow{\textbf{Tor}} \Prov.\Enclave$: \texttt{RA} (Section \ref{teeprel}) 
  \item \HD $\rightarrow$ $\Prov.\Enclave$ : $\langle \idtx,D_w,\hshare :=\shareh \rangle$. 
  \item $\Prov.\Enclave$ : Fetch $\langle \idtx,\pshare:=\{\sharep,c_{sec}\} \rangle$ from \Prov.\db.
  \item $\Prov.\Enclave$ : $k \leftarrow \SReconst(\shareh,\sharep)$.
  \item $\Prov.\Enclave$ : $\secret \leftarrow \Dec(k,c_{\secret})$.
  \item $\Prov.\Enclave$ : $res \leftarrow \WMDet(D_w,\secret)$.
  \item $\Prov.\Enclave \rightarrow \HD$ : $\langle res \rangle $.
  \item $\Prov.\Enclave$: [Erase All]
  \item \HD: \texttt{\textbf{If}} $res \equiv 1$: \textbf{return} \texttt{Valid}\\
        \hspace*{0.5cm} \texttt{\textbf{Else}}: \quad  \textbf{return} \texttt{Invalid}
\end{enumerate}
\end{puppyprotocol}
\subsection{\MPC-based Construction (\puppy-\MPC)}\label{puppy2pc}
\begin{puppyprotocol}[label=verifygc]{\puppy-\MPC Verification}
On input of $\idtx$, $D_w$, and $\hshare$ from \HD, and \db from \Prov, the verification phase is as follows: 

\begin{enumerate}
    \item \HD $\xRightarrow{\textbf{Tor}} \Prov:$ $\idtx$
    \item \Prov: Fetch $\langle \idtx,\pshare \rangle$ from \Prov.\db.
    \item \Prov as a garbler: 
     \begin{itemize}
        \item Compute $(\garbledCircuit,\encodingInfo,\decodingInfo) \leftarrow \Garble(1^\lambda,\verifycircuit)$, where \verifycircuit is the circuit representation of verification, \encodingInfo is the encoding information used to encode an initial input to garbled input, and \decodingInfo is the decoding information that maps garbled output to final output.
        \item Compute $\garbledInput_{\Prov} \leftarrow \Encode(e, \pshare).$
        \item Send $(\garbledCircuit, \decodingInfo, \garbledInput_{\Prov})$ to \HD.
    \end{itemize}
    \item \HD as an evaluator: 
    \begin{itemize}
        \item Compute OTs with \Prov for its input (bits) \hshare to obtain garbled input $\garbledInput_{\HD}$.
        \item Compute $\garbledOutput_{res} \leftarrow \GarbledEval(\garbledCircuit,(\garbledInput_{\HD},\garbledInput_{\Prov}))$ to obtain garbled output $\garbledOutput_{res}$.
        \item Decode $\garbledOutput_{res}$ using $\decodingInfo$ to get verification result $res$ by $res \leftarrow \Decode(\decodingInfo,\garbledOutput_{res})$. 
        \item Return \texttt{Valid} if $res = 1$, or \texttt{Invalid} o.w. 
    \end{itemize}
\end{enumerate}
\end{puppyprotocol}

This section presents another \puppy instance using \MPC. 
\puppy-\MPC does not face the secret size challenge as \puppy-\Enclave faces. 
Therefore, we could omit encrypting the watermarking secret and secret sharing the encryption key.
We can directly secret share the watermarking secret (see Appendix \ref{com2pcefficient}) to have better performance.
Since the modification of ownership is trivial, we describe the verification phase of \puppy-\MPC in this section.

During the verification of \puppy-\MPC, \HD verifies the watermark of \Ow using $D_w$ and $\hshare$ with $\idtx$. 
Recall that \MPC allows the parties to privately run any function (represented by a circuit) without revealing neither their inputs nor any intermediate values to each other. 
To achieve such \MPC, OT and GC are used (see Appendix~\ref{sec:primitives} for their formal definitions).
In \puppy-\MPC, \HD and \Prov agree on a circuit \verifycircuit that computes the steps 5 and 6 in Protocol~\ref{verifyTEE}, i.e., reconstructing \secret from input tokens and performing watermark detection algorithm for an input data $D_w$ using \secret. 
Then, \HD and \Prov run \MPC for that circuit with \HD's input, $D_w$ and $\hshare$, and \Prov's input, $\pshare$. 
\HD initiates the protocol by sending $\idtx$ to \Prov over an anonymous communication channel.\footnote{There are more advanced MPC protocols \cite{anonympc21,anonympcKlingerBM23} that allow parties to hide their identities from each other while securely computing a function. Such MPC design can be deployed to satisfy \secthree without using anonymous communication channels.} 
Once \Prov receives the request, it retrieves \pshare from \db using the received $\idtx$.
\Prov then acts as a garbler in a \MPC computation of the circuit \verifycircuit. 
i.e., \Prov computes a garbled circuit $\garbledCircuit$ of \verifycircuit
and its garbled input $\garbledInput_{\Prov}$ of \pshare, and sends them to \HD with the decoding information $\decodingInfo$.
Then, \HD, as an evaluator of the \MPC, computes OTs to obtain its garbled input $\garbledInput_{\HD}$ of $\hshare$ without revealing \hshare to \Prov, evaluates \garbledCircuit with inputs, $\garbledInput_{\HD}$ and $\garbledInput_{\Prov}$, and obtains the garbled output $\garbledOutput_{res}$.
Finally, \HD acquires the verification result $res$ by decoding $\garbledOutput_{res}$ with $\decodingInfo$ and
returns \texttt{Valid} if $res$ is $1$ (i.e., it detects the watermark of \Ow in $D_w$) or \texttt{Invalid} otherwise. 
Protocol~\ref{verifygc} illustrates this verification phase in \puppy-\MPC. 

\section{Implementation}
We now describe a proof-of-concept (PoC) implementation of \puppy-\Enclave and \puppy-\MPC outlined in Sections~\ref{subsec:puppytee} and ~\ref{puppy2pc}, respectively.
Since the ownership generation phase occurs only once per watermarked \data and its overhead on top of watermark insertion is minimal, we focus on the watermark verification phase.
We use a simple 2-out-of-2 XOR-based secret sharing scheme, AES as an encryption scheme, and a random number for $\idtx$ for our PoC.

\subsection{\puppy-\Enclave Implementation}\label{puppyteeImp}
\textbf{\textit{\HD Communicating with \Prov.}}
\HD conducts remote attestation (RA) with \Prov.\Enclave and establishes a TLS session directly with the \Enclave.
\HD uses this encrypted channel to send necessary data (i.e., $\idtx$, $D_w$, and $s_\HD$) and receive verification results. 

The implementation of RA and TLS session setup between \Prov and \HD is inspired by the SGX-RA-TLS whitepaper~\cite{sgxratls}.
Briefly, the idea is to cryptographically bind the public key of the enclave to the RA certificate (signed by Intel) during the RA procedure.
RA is implemented using the SGX Data Center Attestation Primitive (DCAP)~\cite{sgxdcap}, and RA certificate verification and TLS handshake are implemented using Intel SGX SDK version 2.17.\\
\textbf{\textit{\Prov Running Verification.}}
\Prov consists of an untrusted host application and a \Enclave running the verification code.

\noindent \textbf{$\bullet$ Untrusted Host Application.}
The host application running on \Prov is responsible for: (1) handling connection requests from \HD, (2) retrieving shares from \db upon request from the enclave, and (3) returning the output of the enclave to \HD.
It is implemented in \texttt{C++} and we use SQLite as the \db.

\noindent \textbf{$\bullet$ TEE.}
We used an Intel SGX enclave~\cite{mckeen2013innovative} and the Intel SGX SDK version 2.17 to implement the \Enclave\footnote{Intel has officially announced that their consumer-grade CPUs will stop supporting SGX~\cite{IntelSGXDeprecation}. However, since their server-grade CPUs will continue supporting SGX and we envision \puppy-\Enclave to run on servers, we do not consider this as an issue.}.
The secret reconstruction and watermark detection functionalities are written in \texttt{C++}.
After reconstructing \secret, it is used as input to the \WMDet function that attempts to detect a watermark within \data sent from \HD.
The enclave then encrypts the result of \WMDet with a key pre-shared with \HD (i.e., the TLS session key) and outputs it to the host application.

\subsection{\puppy-\MPC Implementation}
We use \texttt{MP-SPDZ} \cite{mp-spdz}, a general-purpose MPC framework to implement \puppy-\MPC. We run \puppy-\MPC (as a PoC) using Yao's garbled circuit~\cite{yao86}. 2PC protocol based on garbling scheme deployed by \texttt{MP-SPDZ} adopts Half-Gate method~\cite{halfgateZahurRE15} compatible with Free-XOR~\cite{freexorKolesnikovS08} method to decrease the number of garbled gates needed to be sent to the evaluator (\HD) from the garbler (\Prov).
\section{Evaluation}\label{performance}
We now evaluate the security and performance of the \puppy instances.
\subsection{Evaluation of \puppy-\Enclave} \label{subsec:eval_puppy_tee}

\subsubsection{Security Analysis}\label{security}
This section discusses how \puppy-\Enclave satisfies the security requirements specified in Section~\ref{requirement}. 
Due to space limitations, we provide the full security proof in Appendix \ref{teeproof} based on \puppy's formal ideal/real-world security definition presented in Appendix \ref{securitydef}.

\noindent $\bullet$ \textbf{\secone (Watermarking Secret Protection)}. 
The secure authenticated channel prevents any third party from eavesdropping on the messages communicated between \HD and \Prov.\Enclave. 
RA ensures that the correct \texttt{Verify} program is running on \Prov.\Enclave, which isolates the execution of \texttt{Verify} from the rest of \Prov.
In addition, the 2-out-of-2 secret sharing and encryption scheme guarantee that neither \HD nor \Prov learn anything about \secret from their tokens. 
Finally, a secure watermarking scheme ensures that \HD cannot remove or forge the embedded watermark or learn \secret from $D_w$.
Therefore, even if adversary \A corrupts either \HD or \Prov, \puppy-\Enclave satisfies \secone.\\
\noindent $\bullet$ \textbf{\sectwo (Watermarked Asset Protection)}.
Since \Prov.\Enclave and \HD establish a secure channel after RA, no other party (even \Prov) can learn anything about $D_w$. 
Thus, \sectwo is satisfied.\\
\noindent $\bullet$ \textbf{\secthree (Privacy)}. 
Recall that an anonymous channel is used for communication between \HD and \Prov.
Thus, while \Prov learns which $\idtx$ is being queried, it does not learn who queried it.
Therefore, \Prov cannot learn about any transactions between \Ow and \HD, thus satisfying \secthree.

\noindent\textbf{A note on Side-Channel Attacks against TEEs.} 
We acknowledge that there is a large body of research regarding side-channel attacks on TEEs, such as~\cite{brasser2017software,fei2021security,liu2015last,dall2018cachequote,gotzfried2017cache,moghimi2017cachezoom,ControllChannelSideChannel,wang2017leaky,plundervolt,chen2021voltpillager,spectre,meltdown,foreshadow}.
Such attacks are considered out of scope in this paper and \puppy-\Enclave does not actively defend against them.
We refer to prior work~\cite{TSGX,Tamrakar2017,brasser2019drsgx,aexnotify2023constable} for the means to mitigate such attacks as well as \cite{randmets2021overview} for a comprehensive discussion of attack vectors on TEE and countermeasures against such attacks.
\subsubsection{System Analysis}
We now show that \puppy-\Enclave satisfies the system requirements defined in Section~\ref{requirement}. 

\noindent $\bullet$ \sysone (\textbf{Owner-free verification}). 
Since \secret is shared between \HD and \Prov and reconstructed only inside \Prov.\Enclave, the \puppy-\Enclave verification phase does not rely on \Ow.

\noindent $\bullet$ \systwo (\textbf{Unlimited verification}). 
Recall that \Prov and \HD each hold one share of \secret.
For each verification, \secret is securely reconstructed within \Prov.\Enclave, and all \secret-tainted data are afterward erased from within \Prov.\Enclave.
Furthermore, $D_w$ remains the same for each verification.
Therefore, the number of times that the verification can be conducted is unlimited. 

\noindent $\bullet$ \systhree (\textbf{Holder Friendliness).} \HD is only required to engage with \Enclave held by \Prov.
\HD sends its input ($D_w, \hshare$) to \Enclave and receives the output of the computation ($res$).
These communication and computation do not require \HD to perform more expensive (if not negligible) operations than it does in its routine online interaction. 

\subsubsection{Performance Overhead Evaluation}\label{overhead}
To demonstrate the versatility of \puppy-\Enclave, we selected four types of symmetric watermarking schemes from the literature and converted them into publicly verifiable ones: (1) FreqyWM~\cite{isler22}, (2) Shehab et al.~\cite{shehab2007watermarking} (hereafter referred to as ``OBT-WM''), (3) Lou et al.~\cite{lou2007copyright} (hereafter referred to as ``IMG-WM''), and (4) Adi et al.~\cite{adi2018turning} (hereafter referred to as ``DNN-WM''). 
Next, we briefly describe the four watermarking schemes (see Appendix \ref{watermarkingschemes} for the full description) and the reasons for choosing them.

\textbf{FreqyWM.}
This represents a category of watermarking schemes that is capable of inserting a watermark into a data set regardless of data type.
After generating a histogram (based on the frequency of appearance of repeating tokens) from the data set, FreqyWM optimally chooses a pair as a watermarking pair using a secret key $K$, modulus of a secret integer $z$, and an assigned budget. 
The total size of the combined secret values is $\sim 2$ kB.

\textbf{OBT-WM.}
This represents a category of watermarking schemes for numerical databases.
OBT-WM partitions the database based on a secret key $K$, primary keys, and a predefined secret number of partitions $num\_partitions$ and watermarks the partition secret watermark $wm$.
The total size of the combined secret values is approximately 80 bytes.

\textbf{IMG-WM.}
This represents a category of image watermarking schemes.
IMG-WM assumes a monotone image (i.e., pixel values are either $0$ or $255$).
It watermarks a given image by a secret image watermark using Discrete Wavelet Transform (DWT) technique and sharing resulting in two shares as two more secrets.
An open-source implementation of IMG-WM is provided at~\cite{imagewatermark} that uses OpenCV~\cite{opencv} for the detection phase.
The total size of the combined secret values is approximately $240$ kB.

\textbf{DNN-WM.}
This represents a category of machine learning model watermarking schemes.
DNN-WM uses a well-known attack called ``backdooring'' to purposefully train the model to categorize a set of predefined images (called $triggerset$) with a wrong label (called $watermark\_label$).
$triggerset$ and $watermark\_label$ are the secret values for DNN-WM.
An open-source implementation of DNN-WM is provided at~\cite{nnwatermark} that uses PyTorch~\cite{pytorch} for the insertion/detection phases.
The size of the sample $triggerset$ in the implementation is about 17 MB and $watermark\_label$ is 4 kB, therefore the total size of the secret value results in around 17 MB.

As we implemented FreqyWM and OBT-WM in \texttt{C} from scratch, we focused on implementing them for the SGX environment so that they do not depend on third-party libraries that are not supported by SGX (e.g., \texttt{libc}).
On the other hand, IMG-WM and DNN-WM depend on complex libraries which rely on many functionalities that are not provided in SGX (e.g., standard I/O).
To overcome this issue, we used the Gramine library OS~\cite{tsai2017graphine} to run such libraries within the enclave.
Gramine is an open-source project that allows any application to run within the enclave without any modification.
It also enables applications to use external libraries without compromising the security provided by the SGX platform.
Thus, this implies that other types of watermarking schemes that are not described here can be converted into publicly verifiable ones using \puppy. 

\subsubsection{Experimental Setup}\label{eval_setup}
The four watermarking verification schemes ran on an Azure DC8 v2 Confidential Computing VM (8 vcpus, 32 GiB memory, 168 MiB secure memory) running Ubuntu 20.04 LTS OS with kernel version 5.15.0-1020-azure.
FreqyWM and OBT-WM were implemented in \texttt{C/C++} while IMG-WM and DNN-WM were implemented in \texttt{Python} version 3.8.
IMG-WM used OpenCV version 4.6.0.66, and DNN-WM used PyTorch version 1.12.1. 
Both ran on Gramine v.1.2.

We divided the verification phase into five tasks:\\
\noindent $\bullet$ \emph{Establish session:} \Prov and \HD engage in RA (when \Prov is running within the enclave) and establish a TLS session. \\
\noindent $\bullet$ \emph{Receive data:} \Prov receives data from \HD.\\
\noindent $\bullet$\emph{Reconstruct secret:} \Prov reconstructs \secret.\\
\noindent $\bullet$\emph{Detect watermark:} \Prov uses \secret to detect presence of watermark.\\
\noindent $\bullet$ \emph{Terminate session:} \Prov terminates the TLS session with \HD.

We also measured the time to load the watermark secret into PyTorch data structures for DNN-WM detection.
We ran the watermarking detection algorithm for all four schemes inside and outside the enclave to compare the overhead.
To remove network latency we ran \HD and \Prov on the same machine.
The numbers reported are an average of $10$ measurements.

\subsubsection{Evaluation Results}\label{performacetee}

\begin{figure}[htp]
\hspace*{-0.7cm} \includegraphics[width=0.55\textwidth]{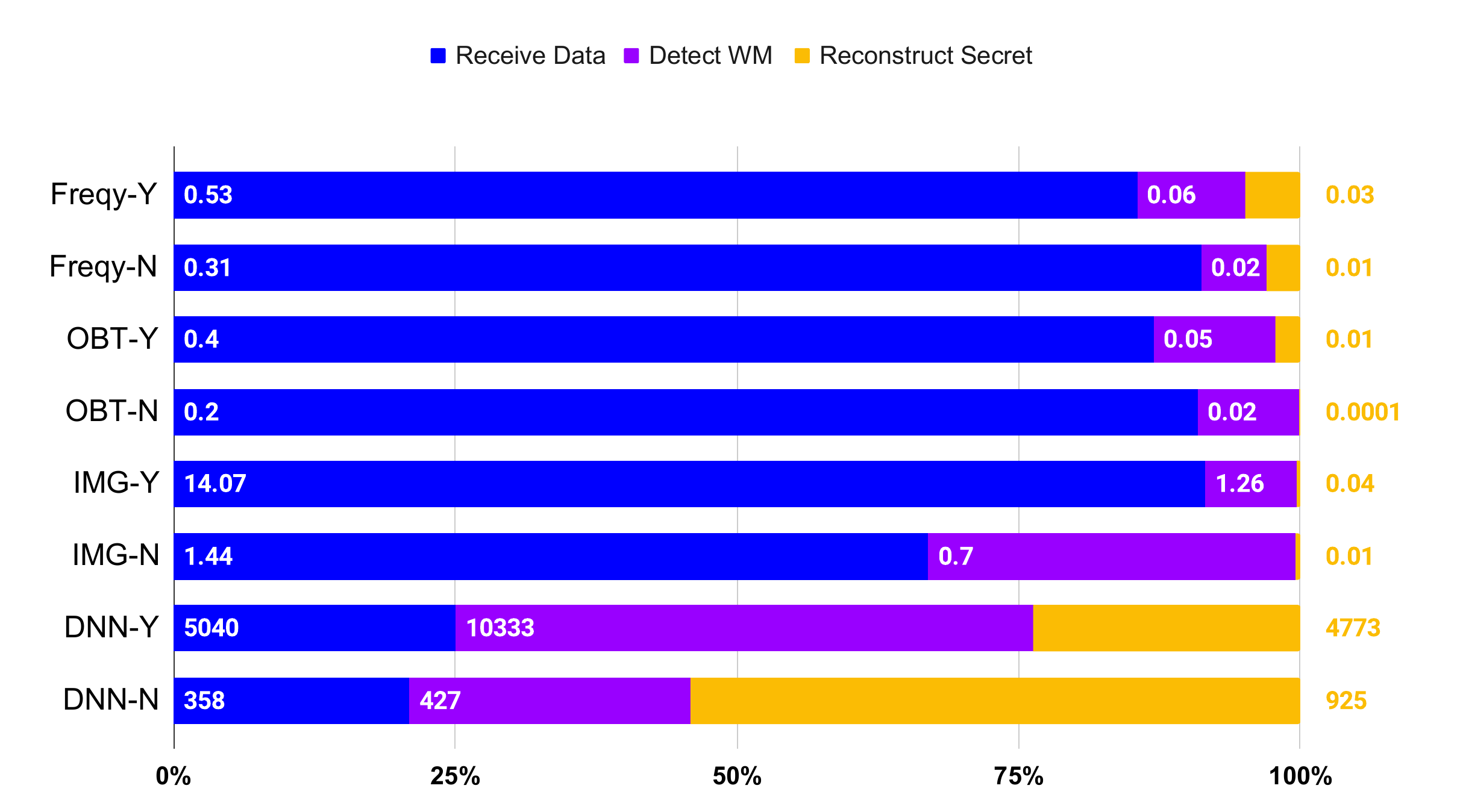}
\caption{Average latency of operations in ms for four watermarking detection schemes. ``Y'' indicates the algorithm running within SGX enclave and ``N'' indicates otherwise.}\label{fig:puppyeval}
\end{figure}

According to our experimental results, the session establishment phase has a relatively high overhead ($\sim6.5$ ms within SGX while $\sim2.25$ ms without SGX) due to the certificate verification. 
This is because \HD has less information to verify when \Prov is not running within the enclave, since there is no RA required.
On the other hand, \HD must verify the RA proof when \Prov is running within the enclave, resulting in additional overhead.
Since the secure channel establishment/termination incurs only a deterministic amount of overhead, we remove it from future measurements.
As for the remaining measurements, ``Receive data'' will indicate the opening and reading of locally stored data into memory.

We run the verification phase of the four watermarking schemes to measure the overall performance of the phase, broken down to the smaller tasks we defined in Section~\ref{eval_setup}.
We also take the measurements twice, once within the SGX enclave (indicated as ``Y'' in the figure) and once outside (indicated as ``N''), to measure the overhead introduced by using a TEE.
Figure~\ref{fig:puppyeval} shows the result of the performance evaluation, which compiles the ratio of each operation relative to the total latency alongside the absolute latency measurement.
We analyze the evaluation results below.

\textbf{FreqyWM and OBT-WM latency.}
Freqy-WM and OBT-WM have smaller secret sizes compare to DNN-WM/IMG-WM.  
To increase its efficiency, we modified the \puppy-\Enclave introducing another secure instance in Appendix~\ref{comefficient}. 
The latency measurements for FreqyWM and OBT-WM detection algorithms in Figure~\ref{fig:puppyeval} shows that each operation incurs (a small amount of) overhead when running inside the enclave.
We can also observe that the ``Reconstruct secret'' phases in the Freqy-WM and OBT-WM are extremely fast outside of the enclave but not so within the enclave.
This is because of the small overhead in entering and exiting the enclave and this overhead dominates the measurement.

\textbf{IMG-WM latency.}
From Figure \ref{fig:puppyeval}, we can observe that the ``Receive data'' latency is higher when loading into secure memory, compared to loading into regular memory.
This is because accessing the shared image file outside the enclave requires Gramine to securely copy the data into the enclave, resulting in the en/decryption of the file, which is expensive. 
Other operations show no significant difference in terms of latency when running inside or outside the enclave.
Overall, the total process is in the order of milliseconds, which is acceptable.

\textbf{DNN-WM latency.}
Unlike the other schemes, the size of the secret used to generate the DNN model watermark is large; $17$ MB for the sample trigger set holding $100$ images.
Since each image is large, secret-sharing these images takes several seconds.
However, we have designed \puppy in such a way that this is not an issue (see Section~\ref{design} for more information).

We observe from Figure~\ref{fig:puppyeval} that all operations of DNN-WM are one to two orders of magnitude slower when running inside the enclave.
This is because DNN-WM requires a large amount of data to run the detection algorithm; for instance, the PyTorch python library itself is several hundreds of MB.
This exceeds the total usable secure memory for SGX enclave supported by the CPUs used in DCs v2 VMs, which is around 100 MB.
Since the required enclave size is larger than the usable secure memory size, SGX requires enclave pages to be swapped and en/decrypted, which results in the high latency we see in Figure \ref{fig:puppyeval}.\footnote{The only way to overcome this limitation is to use machines equipped with Intel Icelake CPUs (e.g., DCsv3 and DCdsv3 Azure Confidential Computing VMs), although such machines are scarce and are hard to obtain.}

\noindent\textbf{Note:} We acknowledge that introducing Gramine, image processing, and machine learning libraries into \Enclave increases attack surface. 
This is a limitation of \puppy-\Enclave and a trade-off between versatility and security.

\subsection{Evaluation of \puppy-\MPC}\label{2pceval}
\subsubsection{Security Analysis}\label{2pcsecurity}

We now provide a brief security analysis of \puppy-\MPC considering the requirements in Section~\ref{requirement}.
We omit the formal proof of \puppy-\MPC since it follows a very similar approach to \puppy-\Enclave's. 
    \noindent $\bullet$ \textbf{\secone (Watermarking Secret Protection)}: Secret sharing guarantees that both \HD and \Prov cannot learn anything about \secret from their shares. 
    During the verification phase, given that all underlying primitives, GC and OT, are secure (see Appendix~\ref{sec:primitives} for definitions), \MPC guarantees that \HD learns nothing about \Prov's input or intermediate results, other than $res$, and \Prov learns nothing about \HD's input or intermediate results. Hence, \secone is satisfied.\\
    \noindent $\bullet$\textbf{\sectwo (Watermarked Asset Protection)}: The watermarked \data $D_w$ stays with \HD and OT ensures that $D_w$ is not revealed to \Prov. Thus, \sectwo is satisfied.\\
    \noindent $\bullet$ \textbf{\secthree (Privacy)}: \Prov cannot identify \HD, as they communicate over the anonymous communication channel. Thus, \secthree is satisfied.

\subsubsection{System Analysis}\label{2pcsystem} 
\puppy-\MPC satisfies \sysone and \systwo similar to the system analysis of \puppy-\Enclave in Section~\ref{subsec:eval_puppy_tee}. 
Our performance analysis in the following section shows that \puppy-\MPC protocol introduces ``reasonable'' overhead to \HD, thus satisfying \systhree. 

\subsubsection{Performance Analysis of \puppy-\MPC}\label{2pcperform}
We use the \texttt{MP-SPDZ} \cite{mp-spdz} framework to evaluate the efficiency of \puppy-\MPC. 
We set a baseline as the case where all computations during verification are computed without \emph{any} secure computation (equivalent to the \texttt{No SGX} case in \puppy-\Enclave).
Based on this baseline, we evaluate the computation overhead of \puppy-\MPC on \HD and \Prov. 
We first measure the most efficient watermarking scheme, FreqyWM, out of four watermarking schemes in Section~\ref{overhead}, using \puppy-\MPC and compare it against the most expensive watermarking scheme, DNN-WM, implemented in \puppy-\Enclave.
We used the same parameter settings for FreqyWM as in Section~\ref{performacetee} on the same machine.

\textit{Optimizations.} To improve the performance of \puppy-\MPC, we introduce a few modifications to the detection algorithm of FreqyWM (see Appendix~\ref{freqywmdetect} for the modified algorithm). 
We analyzed the detection algorithm to identify the steps that are not required to be computed via 2PC without violating the underlying security, thus improving performance.
Through our analysis, we have identified that the first step (creating a histogram of $D_w$ -- arithmetic) and the last step (checking if the number of detected watermarks is over a given threshold -- boolean) in the algorithm do not need to be computed via 2PC. 

\textit{Evaluation results.} 
The global data sent (e.g., the garbled circuit) during the 2PC is around $1.9$ GB. Compared to the amount of data sent in the \puppy-\Enclave instance which is in the order of KB, this is orders of magnitude larger and should be taken into consideration when deploying \puppy-\MPC. 
We can see that the time to verify a watermarked \data is $\sim$16 seconds, which is around $\times160$K slower compared to when running FreqyWM via \puppy-\Enclave ($\sim$0.1ms).
Furthermore, the execution time of FreqyWM via \puppy-\MPC is longer than the execution time of DNN-WM via \puppy-\Enclave ($\sim$15 seconds) which is by far the most expensive and complex watermarking technique.
Therefore, since the other watermarking schemes are more expensive and complex than FreqyWM, we estimate that running them on \puppy-\MPC will take longer than the result shown above, thus will be more expensive than any measurements taken using \puppy-\Enclave.

Overall, our evaluation results show that \puppy-\Enclave is more efficient than \puppy-\MPC supporting the observations by prior works such as~\cite{privsignal22}.
We identify implementing \puppy using more efficient \MPC (e.g., ones listed in \cite{hastings2019sok}), FHE, and FE, while optimizing circuit complexity across different watermarking techniques and balancing holder workload as a challenge.
Due to the limited amount of space, we consider the above a potential pathway for future work.
\section{Optimization of \puppy via Memoization}\label{memoization}

Consider a scenario in which two data marketplaces that act as holders, say $\HD_1$ and $\HD_2$, happen to verify two different assets that resemble some asset that they both possess.\footnote{Note that as mentioned in Sections~\ref{intro} and \ref{wmprimer}, watermarking schemes are expected to be robust against any (tolerable) modification since it is very common to assume that an asset may be tampered (un)intentionally.}
Naturally, both holders will independently execute the verification phase with \Prov for the same watermarking secret while their assets are insignificantly different.
Such repetitive computations lead to communication and computation overhead for both \HD and \Prov especially when there are a large number of holders.
Moreover, a (malicious) \HD can easily flood \Prov with requests, conducting a denial-of-service attack.
Overall, a \puppy instance has to address two challenges: 1) \textit{Redundancy} due to repetitive verification requests; and 2) \textit{Denial-of-Service} (attack) both of which can congest \Prov.

To overcome the aforementioned challenges, we introduce a \textit{memoization} mechanism to \puppy via \textit{caching}. 
In this approach, before \HD starts the verification phase of an instance (e.g., \puppy-\Enclave), it first checks with \Prov if a similar \data and $\idtx_i$, which it is willing to verify, has been previously requested by \textit{anyone}. 
To enable this functionality, \Prov caches the verification requests (and their verification results), and if a verification request results in a cache hit, then \Prov serves the request from the cache. 
By doing so, both \Prov and \HD avoid redundant computations. 

There are two important aspects to consider in any caching problem: 1) \textit{caching decision}; and 2) \textit{cache replacement decision}. 
During \textit{caching decision}, whenever a new entry is requested, \Prov has to decide whether the entry should be stored in the cache or discarded.
After deciding if the new entry will be stored, the cache, which has a limited capacity, has to be updated. 
Thus, if the capacity is exceeded, then \Prov evicts an entry in its cache by running a \textit{cache replacement algorithm}. 
We base our approach on the well-known least-recently-used (LRU)~\cite{lruBianchiDCB13,lruhitGarettoLM16} rule due to its simplicity and ability to handle time-correlated requests efficiently~\cite{LaoutarisCS06} and adapt it to the peculiarities of watermark-verification caching.

\noindent \textbf{Intuition:}
For \puppy, we identified two opportunities to improve the caching decision and the eviction policy. 

\emph{Caching Decision.}
Watermarking detection algorithms 
succeed as long as an asset has not been modified more than some tolerance threshold \cite{isler22}.  
If the similarity of two assets is above a threshold, they will be considered as the same asset entry. 
However, in a traditional caching decision, two assets, even if they differ by a few bits, would be considered different and cached independently. 
To prevent the creation of such caching entries, we aim to avoid different sub-samples or partitions of an asset.
Since we do not intend to cache the actual asset, our focus is on caching the verification result regarding whether a watermark was present in a given asset. 
Thus, we use perceptual hashing \PHash (see Appendix~\ref{lsh} for its formal definition) to check if two assets are similar enough to be considered as the same asset.

\emph{Eviction Policy.}
If \puppy decides to cache an entry although there is no space in the cache, an entry from the cache must be evicted by running \emph{cache replacement (eviction) policy}.
LRU, the cache replacement policy we base on, uses the recency of requests indicating that the most recently used are more valuable and shall be left in the cache.
We have identified that replacing cache entries in \puppy must take into consideration the following two additional parameters: 1) similarity between the requested and cached item; and 2) whether the cached verification result was ``0'' or ``1''. 

To better illustrate this, consider FreqyWM\footnote{Similar observations can be made to OBT-WM.} as an example.
In FreqyWM, if an asset $D_w$ is similar to another asset $D^i_w$ with $70\%$, and the result of verification on $D_w$ is $1$ then if there is another asset $D^j_w$ similar to $D^i_w$ with $90\%$, the verification result on $D^j_w$ is highly likely $1$ given the analysis and the results provided in the FreqyWM paper.
Thus, our approach shall favor caching requests having a lower similarity but a positive (1) verification, since they have more utility by being relevant to more future requests with higher similarity values (a cached $70\%$ positive result allows a better guess for a future $80\%$ request than a positive cached result with $90\%$ similarity).
The opposite applies to negative (0) verification results -- in this case, we give preference to keeping items of higher similarity in the cache. 

Note that introducing this mechanism requires the following modifications to our verification protocol: 1) \HD computes a perceptual hashing and similarity; and 2) \Prov receives the result of the verification, perceptual hash value, and similarity result. 
However, the overhead introduced to \HD and \Prov is insignificant, and the approach does not harm the security of \puppy (see Appendix \ref{caching_security} for further discussion).

\begin{puppyprotocol}[label=verify_cache]{Verification w/ Memoization}
On input of $\idtx_i$, $D'_w$, and $D^i_w$ from \HD, and \cache from \Prov, the verification phase is as follows: 
\begin{enumerate} 
  \item \HD: $h' = \PHash(\hashparam,D^i_w,D'_w)$ 
  \item \HD: $sim' = \Sim(D^i_w,D'_w)$
  \item \HD $\xRightarrow{\textbf{Tor}}$ \Prov : $\langle \idtx_i, h', sim'\rangle$. 
  \item \Prov : $\langle res_r, sim_r \rangle = \cache.\Get(h',\idtx_i)$ 
  \item \Prov : \\
   \hspace*{0.25cm} \texttt{\textbf{If}} $res_r \equiv 1 \land sim_r \geq sim' $: \textbf{return} $res_r$\\
   \hspace*{0.25cm} \texttt{\textbf{ElseIf}} $res_r \equiv 0 \land sim_r \leq sim'$: \textbf{return} $res_r$ \\
   \hspace*{0.25cm} \texttt{\textbf{Else}}: \quad  \textbf{return} $res_r=\texttt{None/Invalid}$
   \item \Prov $\xRightarrow{\textbf{Tor}}$ \HD : $res_r$
   \item \HD: \texttt{\textbf{If}} $res_r \equiv \texttt{None/Invalid}:$ \texttt{Compute} \texttt{Verify} protocol w/ \Prov.
\end{enumerate}
\end{puppyprotocol}
\subsection{Memoization via Proportional Caching}\label{proportionalcache}

Next, we define our approach in more detail. 
Let $\mathcal{C}$ be a cache with size $c$ supporting the following two functions: 1) \Get for searching for a request in \cache; and 2) \Put inserting an entry into \cache according to the cache decision policy and (if needed) updating \cache according to the eviction policy. 
\HD first computes a perceptual hash $h'$ of $D'_w$ and $D^i_w$ associated with $\idtx_i$ that \HD wishes to run verification protocol and a value of similarity $sim'=\Sim(D'_w,D^i_w)$.\footnote{Any choice of similarity can be integrated, e.g., Cosine, Jaccard.} 
\HD sends $\langle h', \idtx_i,sim'\rangle$ to \Prov. 
\Prov checks $C$ to determine if $\langle h',\idtx_i\rangle \in C$. 
If it exists, it retrieves $res'$ and the similarity stored $sim_r$ for the given entry from $C$. 
If $res_r \equiv 1 \land sim_r \geq sim'$ or $res_r \equiv 0 \land sim_r \leq sim'$, \Prov returns $res_r$ to \HD.\footnote{While we refer $res$ as a binary value, note that a proof of the verification, i.e. digital signature, may need to be stored at \cache and shared with \HD. However, for simplicity, we do not include it in our representation.}  
Otherwise, \Prov returns not available (e.g., \texttt{NONE}) which means that \HD has to perform the verification protocol by running a \puppy instance (see Protocol \ref{verify_cache} for the verification phase with memoization).  
After a cache hit, the entry in \cache is moved closer to the top of the cache depending on its similarity as explained next. 

\emph{Caching Decision.} 
Assume \Prov receives a verification results $\langle h_l,id_l,res_l,sim_l \rangle$. 
\Prov checks if the similarity is above a threshold $t$, and then adds it to \cache to decrease the false positive ratio.
If $res_l = 0$, \Prov computes $f_0(sim_l)$ as defined by Equation~\ref{eq:res_zero} to determine the position of the entry in \cache.
Otherwise, meaning that $res_l = 1$, \Prov computes $f_1(sim_l)$ as defined by Equation~\ref{eq:res_one} (see Appendix~\ref{intervalshift} for the derivation of the equations). 
After calculating the position, $p_l$, of the entry, it is inserted to \cache at $p_l$ if $sim_l \geq t$ if \cache has space. 
If \cache does not have enough space, the eviction procedure is computed. 

\begin{equation}\label{eq:res_one}
    f_1(sim)= \frac{(c-1)\times sim}{100}
\end{equation}
\begin{equation}\label{eq:res_zero}
    f_0(sim)= c- \frac{(c-1)\times sim }{100}
\end{equation}
\emph{Eviction procedure.} As in LRU, the last entry in \cache is removed since the last entry has the lowest priority to accommodate the new entry. 
However, if $\langle h_l,id_l \rangle$ exists, then $sim_l$ and the similarity $sim'$ in \cache are compared by considering the value of $res_l$. 
As discussed previously, for $res_l=0$, \Prov stores the entry having a higher similarity while storing the entry having a lower similarity for $res_l=1$.

\textbf{\emph{Cache Hit Ratio.}} 
We now analyze the cache hit (HT) which describes the situation when the request of \HD is successfully served from the cache. 
Let us now define the cache hit ratio (HR)~\cite{lruhitGarettoLM16} as:
\begin{equation*}
    HR = \frac{HT}{R}
\end{equation*}
where $R$ is the total number of requests.
HR depends on various factors: 1) watermarking schemes; 2) similarity metrics and threshold; 3) perceptual hashing parameters; 4) the total number of assets and their probability distributions. 
We briefly show how these metrics affect HR and leave a more detailed examination for Appendix~\ref{furthermemoization} due to page limitations. 

\noindent\textbf{Note:}
Many caching strategies must also take diverse entry sizes into consideration and optimize their memory accordingly.
However, since the cached values in \puppy have a fixed size, we do not conduct any optimization via memory management.

\subsection{Evaluation}\label{cacheExpAnalysis} 
To demonstrate its potential, we implemented the memoization approach with various data samples, cache capacity values, and similarity values to analyze the cache hit ratio (see Appendix~\ref{cacheAnalysis} for our experimental setup and results in detail). 
For our experiments, we use the similarity threshold $t$ (i.e., $70\%$) and data samples derived by FreqyWM and implement the proposed cache with different capacities.
In summary, our experimental results show that our proportional-caching approach is exceptionally more effective than the simple LRU implementation for our use case, achieving $60\times$ more cache hit ratio.

\subsection{Further Discussion and Limitations}\label{memoizationlimit}
Although our approach significantly improves the query response time, it has a few limitations: 1) \textit{cache poisoning}; and 2) \textit{false positive} results.
Cache poisoning can be overcome by increasing the threshold value and enforcing \HD to prove that the perceptual hashing is computed correctly (e.g., via ZKP).
False positive (i.e., announcing an asset is watermarked by a certain secret/owner while it is not) ratio is orthogonal to the chosen parameters as discussed in Appendix~\ref{furthermemoization}.
Although our caching approach is simple yet very specific to \puppy, applying our approach to a more complicated watermarking technique (i.e., DNN-WM) and investigating other sophisticated caching mechanisms (e.g., \cite{lhd18,cachelib20}) are interesting future directions we plan to explore.

\section{Discussion} \label{further}
\subsection{One-per-Holder vs. One-for-all-Holders}\label{oneforall}
\puppy does not restrict how \Ow generates watermarked \data (as mentioned in Section~\ref{subsec:system_design}) and supports cases where \Ow shares/sells its \data to multiple \textit{Holders}, in two ways:
\begin{itemize}
    \item \textbf{[One-per-Holder]:} \Ow generates a different watermarked 
    version of the same \data per holder.
    \item \textbf{[One-for-All]:} \Ow generates only one watermarked version of 
    \data and share it with all holders.
\end{itemize}
Note that in the former, the watermarked \data can be used as a \textit{fingerprint} to identify holders who illegally distribute their watermarked \data. 

\subsection{Extending \puppy to Multiple Provers}\label{extendprover}
Security and reliability of \puppy can be enhanced by introducing multiple \textit{Provers}, $\Prov_1,\dots, \Prov_{\np}$. 
To have a more reliable design with the same security guarantee, \Ow encrypts \secret with $k$ and secret shares $k$ with each $\Prov_i$ and \HD, generating $\psharei=\langle \pishare, c_{\secret}\rangle$ and $\hshare$. \Ow sends $\psharei$ to each $\Prov_i$ and $\hshare$ with \HD, such that \HD and $\Prov_i$ can reconstruct the secret. As long as one $\Prov_i$ is available, \HD can run the verification.

On the other hand, security can be improved by \Ow generating a secret sharing of $k$ with $\SGen(k,\{\HD,\Prov_1,\dots, \Prov_{\np}\})$ and distributing the shares according to the well-defined authorization rule, i.e., the access structure. 
The access structure must always include \HD as a party in reconstructing \secret in order to prevent \textit{Provers} from colluding.
One simple way is to use threshold secret sharing (TSS)~\cite{shamir1979share}.
By doing so, the probability of an adversary obtaining the secret information becomes much lower than the case with one \textit{Prover}, since the adversary now has to corrupt threshold-many provers rather than only one.

\subsection{Using More Advanced Secret Sharing Schemes}\label{ssdiscussion} 
Secret sharing (SS) is one of the fundamental building blocks. 
\puppy is agnostic with respect to the underlying SS scheme. It can use threshold 
SS (e.g., 2-out-of-2 used in the current implementation), verifiable SS 
(VSS)~\cite{benaloh1986secret,chor1985verifiable,feldman1987practical,jhanwar2014paillier}, and access controlled SS (ACSS)~\cite{benaloh1990generalized}. 
For example, with VSS, each party can verify if the shares are correctly shared and detect any malicious (i.e., arbitrary) shares. 
With ACSS, \Ow can make sure that \HD communicates with a prover (or a subset of thereof) to verify ownership. 
\subsection{Watermark Selection \& \puppy Integration}
The unlimited verification requirement (\systwo) of \puppy might affect the underlying watermarking schemes.
For example, suppose that we use a secure/robust database watermarking scheme that requires \Ow's involvement during the watermark verification phase. 
Since \puppy frees \Ow from the verification phase, a malicious \HD may keep modifying the database and check with \Prov whether the modified database still contains the watermark by running the verification phase multiple times.
The original watermarking scheme prevents this because, after a certain number of verifications, \Ow can reject the request from \HD.

We emphasize that this attack is highly dependent on the underlying watermarking scheme and may not be applicable to all symmetric watermarking algorithms. 
This requires further investigation.
Also, this attack can be mitigated via rate limiting of the verification phase, i.e., \Ow (or \Prov) setting a limit on the number of verifications of the same \data within a certain time period.  

\section{Related Work} \label{relatedwork} 
\puppy is the first formally defined framework for converting (any) symmetric watermarking to a publicly verifiable one. 
However, there exists some literature proposing specific watermarking techniques with public verification functionality as we provide an overview of them in this section.

Existing publicly verifiable watermarking solutions can be categorized into: 1) \emph{Multi-Watermarks-based}, 2) \emph{Certification (TTP)-based}, and 3) \emph{Zero-Knowledge Proof~(ZKP)-based} (a.k.a. zero knowledge watermarking).
The \emph{multi-watermarks-based} approach generates multiple watermarks on one \data and discloses one watermark for each verification.
On the other hand, the \emph{TTP-based} one assumes a trusted third party (TTP) and announces the watermarking secret(s) with a TTP's approval (e.g., via a certificate). 
Lastly, the \emph{ZKP-based} solutions use ZKPs to allow owners to prove their ownership by sending some additional information. 
Lach et al.~\cite{lach1999robust} proposed a watermarking technique for \emph{Field-Programmable Gate Array} (FPGA) by embedding a large number of watermarks where each watermark is small. 
For each verification request, one watermark is revealed for public verification. 
However, this solution supports only a limited number of verifications. 
In addition, it requires the owner to be online during the verification to reveal the watermark secret for each verification.

Li et al.~\cite{li2006publicly} proposed a publicly verifiable database watermarking where a database owner watermarks its data using its secret key and creates a certificate on the key with the help of a certificate authority (CA). 
The certificate is enclosed with the watermarked database as an \data and any party receiving both the certificate and database can verify the ownership by verifying the certificate, i.e., the signature of the CA on the secret key. 
However, this requires the CA to be trusted by all entities and more importantly, reveals the secret key allowing anyone to remove the watermark. 

Adi et al.~\cite{adi2018turning} proposed two solutions for (deep) neural network watermarking with public verification.
They use backdoor information (e.g., a set of images are incorrectly labeled on purpose) as a watermarking secret and they prove the ownership by showing the knowledge of this information.
One of their solutions works similarly to Lach et al.~\cite{lach1999robust} and has the same limitation (only fixed times of verification available). 
The other solution adds the ZKP process to verify the ownership an unlimited amount of times, and a few other papers~\cite{adelsbach2003watermark,adelsbach2001zero,saha2011secure} also use/build zero-knowledge watermarking protocols.
Watermarking schemes integrating ZKP allow an owner to prove ownership to a third party but face \emph{poor scalability}.
Since the rightful owner must generate a ZKP over the modified \data, it is easy to imagine how this can quickly become overwhelming for the owner, especially considering the fact that they can own multiple \data{s}.
Furthermore, ZKP-based schemes require owners to be online during the watermark detection phase since the owner is the only entity that can produce the proof.
This introduces delay in \data processing and additional communication overhead (see Section~\ref{subsec:challenges}).

\section{Future Work and Conclusion}\label{futurework}
We proposed \puppy, the first formally defined framework that converts any symmetric watermarking to be publicly verifiable without requiring the intervention of \data owners for verification. 
We derived various instances securely realizing our proposed publicly verifiable watermarking framework functionality using cryptographic primitives such as FHE, MPC, and FE in a comparative manner. 
We designed and implemented \puppy-\Enclave and \puppy-\MPC, two secure instances that satisfy all system and security requirements for \puppy.
By implementing four types of watermarking, we showed that \puppy can be practically deployed for various watermarking schemes without modifying them.

\puppy assumes an honest owner(s) during the ownership generation phase; however, a malicious owner can make the whole verification meaningless by distributing wrong tokens or wrong watermarked asset. One idea to avoid this is to enforce the owner to send a proof for every message generated during the ownership generation phase; however, it requires further investigation in the future.
Furthermore, we plan to investigate integration of \puppy to existing Buyer-Seller Watermarking (BSW) ~\cite{chang2010efficient,frattolillo2016buyer,frattolillo2021blockchain,ju2002anonymous,katzenbeisser2008buyer,rial2010privacy,rial2010provably,williams2010importance} and validate asset authenticity (e.g., DECO~\cite{zhang2020deco}, TLS-Notary~\cite{tlsnotary}).

\section*{Acknowledgements}
Devriş İşler was supported by the European Union’s HORIZON project DataBri-X (101070069). 
Nikolaos Laoutaris was supported by the MLEDGE project (REGAGE22e00052829516), funded by the Ministry of Economic Affairs and Digital Transformation and the European Union-NextGenerationEU/PRTR. 
Yoshimichi Nakatsuka was supported in part by The Nakajima Foundation.

\bibliographystyle{plain}

\appendices
\section{Optimization of \puppy via Memoization (Cont'd)}\label{furthermemoization}

\textbf{\emph{Cache Hit Ratio.}}
Before delving deeper into the cache hit ratio analysis, we shall investigate the probabilities of corner cases of our approach as follows: \\
\noindent $\bullet$ $Pr[h_l=h'|sim_l=0]$. Consider a new entry and existing entry in \cache produce the same hashing value but the similarity is $0$. If $res_l=1$, then this entry must be inserted to the top of \cache. 
However, the probability of this happening is closed to zero due to watermarking detection. 
Watermarking techniques such as FreqyWM show that two assets must be similar enough to have a detection resulting a positive value (i.e., $res_l=1$). \\
\noindent $\bullet$ $Pr[h_l \neq h'|sim_l=100]$. 
If two assets are the same, then their perceptual hashing values must be the same.
Thus, the probability of this edge case is zero. 

Therefore, we are interested in probability of two assets with the same perceptual hashing where the similarity of them is over the threshold $t$ denoted as $p_l=Pr[h_l=h'|sim_l\geq t]$. 
$p_l$ determines the probability of $D^l_w$ to produce the same LSH value with $D'_w$. 
$p_l$ can follow different probability distributions (e.g., Uniform, Normal). 
While we leave the concrete and extensive theoretical analysis of the hit ratio of our approach in the future, we refer readers to other caching mechanisms \cite{cacheKing71a,cacheYan022}.

\subsection{Evaluation}\label{cacheAnalysis} 
A caching strategy may need to optimize its memory due to diverse entry size. 
However, in \puppy, values to be cached are fixed and have the same size. 
Therefore, our caching approach has an advantage compared to simple LRU-based caching mechanism thanks to unique features of \puppy; thus, we do not elongate our discussion with a cache memory management. 
We implemented the proportional-caching as defined in Section \ref{memoization} as a dictionary where $\langle h,id\rangle$ is key and $\langle res, sim \rangle$ is value and \texttt{MinHash} as a perceptual hashing.  
In our experiments, we randomly generated $250$ key and value pairs and for each sampled key-pair, we created a cache with various capacity as $[10, 20, 50, 100, 250]$. 
We run our approach $100$ times and take the mean of total computations.

For our evaluation, we compare the proportional caching, denoted by \lruprop which is a modified LRU as defined in Section \ref{memoization}. 
We first implemented a simple LRU denoted by \lrubase. 
In \lrubase, a cache hit occurs if and only if a request $\langle h,id,sim \rangle$ exists in the cache, then it is moved to the head of the cache. 
In \lrubase, there is no similarity check to avoid multiple entries of the same $\langle h,id\rangle$ tuple with different similarity values $sim$. 
To show how our similarity approach affects cache hit ratio if a traditional LRU, we tested two approaches with the following two types of requests: \\
\noindent $\bullet$ \textbf{\textit{Type-1} (Fixed Similarity):} Created $250$ key-value pairs are inserted into a cache according to the caching decision policy. Then the keys ($(h,id)$) are requested with the same similarity value from the cache. \\
\noindent $\bullet$ \textbf{\textit{Type-2} (Flexible Similarity):} Generation of key-value pairs and insertion are the same as \textit{Type-1}. However, when we request a result for a key with some certain similarity, for each key, we randomly assigned different similarity. 
The reason is to evaluate how our similarity check addition we introduce improves the performance.

\begin{figure}[htp]
 \includegraphics[width=0.5\textwidth]{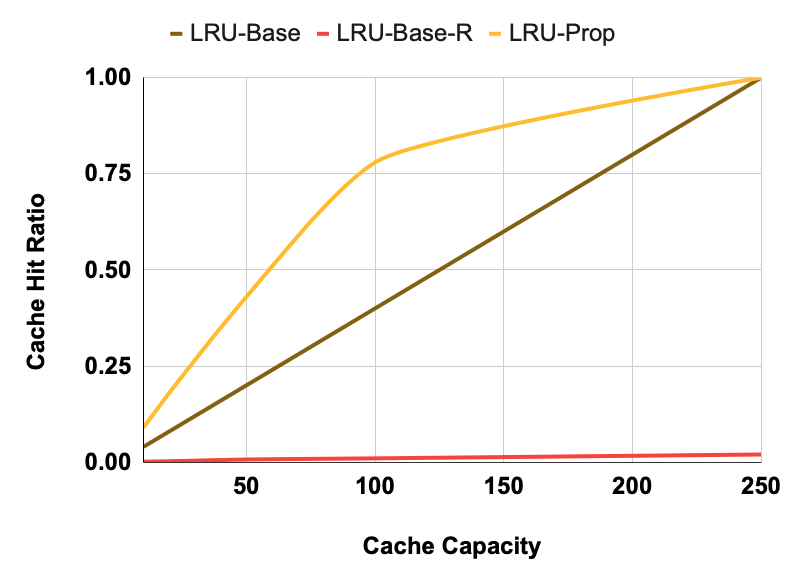}
\caption{Hit ratio analysis of \lrubase and \lruprop.}
\label{fig:lrucompares}
\end{figure}

Figure \ref{fig:lrucompares} illustrates the correlation between cache hit ratio and cache capacity for LRU, and \lruprop. 
\lrubase represents the evaluation results of the cache hit ratio of the traditional LRU using the requests defined as \textit{Type-1} while \lrusim represents the results of LRU when it is computed on the requests defined as \textit{Type-2}.
For \lruprop, there is no significant difference whether the request is \textit{Type-1} or \textit{Type-2}. Thus, there is only one evaluation results for it in the figure.

\textbf{[\lrubase vs. \lrusim].}
As shown by Figure \ref{fig:lrucompares}, if a traditional LRU was directly applied to \textit{Type-2} requests as in our use case scenario, the cache hit ratio of \lrusim is below $0.02$ in average. This is expected since there is no similarity comparison (i.e., returning $res$ by comparing the cached similarity and requested similarity).
However, to have a fair comparison, we evaluated \textit{Type-1} requests, again it assumes that the same similarity values will be requested (which is not realistic real-life scenario).
In this case, \lrubase performs better than \lrusim. 
Since we requested exactly the same entries again, when the cache size is equal to the number of requests (i.e., $250$), \lrubase's hit ratio reaches to $1$.\\
\textbf{[\lruprop vs. \lrusim].}
Let us now compare how \lrusim and \lruprop perform when a real-world use case scenario occurs. 
As shown, \lruprop outperforms \lrusim drastically by $60\times$ in average while it is $1.5\times$ in average compared to \lrubase. 
Overall, our experimental results support our claim that our caching approach is a more effective policy for the real-world use case of data economy design, and a direct application of the LRU caching to \puppy does not improve the performance. 
However, we plan to investigate our approach using real-life assets and fully implement it in large scale (e.g., a higher number of datasets, cache capacities) in the future. 

\subsection{Further Analysis}\label{caching_security}
In this section we further analyse the overhead introduced to \HD in terms of computation complexity and the security analysis when integrating memoization to \puppy. 

\textbf{Computation Overhead.} \HD has to compute a perceptual hashing (e.g., MinHash) per request. 
Our results show that running a $MinHash$ for two (small size) datasets which generated as in FreqyWM, it takes less than $0.01$ms.\\

\textbf{Security Analysis}. Since verification with memoization neither requires any token information or computes any function on the watermarking secret, there is no attack scenario that a malicious \HD or \Prov can break any of the security requirements (see also our discussion in Section \ref{memoizationlimit} for other possible attack scenarios). 

\section{Further Primitives} \label{sec:primitives}  
\subsection{Finding the Position of an Entry}\label{intervalshift}
To determine the position of an entry $\langle h,id,res,sim \rangle$ to add in the cache and move up to the correct position after a cache hit, we shall convert a given similarity $sim \in [0,100]$ to a position in the cache \cache. 
Let us define this conversion $f_{res}(sim)$ considering the value of $res$. 
\begin{equation*}
f_{res}(sim)=
\begin{cases}      
 [0,c-1], \text{ for $res=1$}\\
 c-f_1(sim), \text{ for $res=0$}
\end{cases}
\end{equation*}
We derive our equations for $f$ basing on the an interval shifting $[a,b]\rightarrow [x,y]$ which is formulated as : 
\begin{equation*}
    f(m)= x + (\frac{y-x}{b-a})\times (m-a)
\end{equation*} where $m\in[a,b]$. 
Replacing $[a,b]\rightarrow [x,y]$ with $[0,100] \rightarrow [0,c-1]$, we derive the following equation:
\begin{equation*}
    f(sim)= \frac{(c-1)\times sim}{100}
\end{equation*} where $sim\in[0,100]$. 
Hence applying the equation above, we have the following equations to determine the position of an entry considering its similarity and result in \cache as: 
\begin{equation*}
f_{res}(sim)=
\begin{cases}      
 \frac{(c-1)\times sim}{100}, \text{ for $res=1$}\\ \\
 c- \frac{(c-1)\times sim }{100}, \text{ for $res=0$}
\end{cases}
\end{equation*}

Note that with this conversion, a position of a similarity value (which could be a decimal value) may have the same position with another similarity. 
For instance, let $sim=98$ and $sim=98.5$ be two similarity values of two entries to be cached with the same result and assume $c=100$. 
Then, the positions of the entries will be same.

\subsection{Watermarking Primer (Cont'd.)}\label{wmfurther} 
We define common properties of watermarking which we are inspired from \cite{adi2018turning,barak2001possibility} as follows:
\begin{definition}[Correctness]
The algorithms (\WMKeyGen, \WMGen, \WMDet) should work correctly for an honest owner. In other words, correctness states that an honest owner should verify the watermark with its correct corresponding \secret that is generated honestly.
 Correctness formally requires: 
  \begin{align*}
   Pr[\secret \leftarrow \WMKeyGen(1^\lambda); D_w \leftarrow \WMGen(D_o,\secret): \\ \WMDet(D_w,\secret)=accept]=1
\end{align*}
\end{definition}
\begin{definition}[Unremovability]
An adversary is unable to remove a watermark, even if it knows about the existence of a watermark and knows the algorithm that was used in the process. 
For every PPT algorithm \A, the chance of winning the following game is negligible:
\begin{compactenum}
    \item Compute $\secret \leftarrow \WMKeyGen(1^\lambda)$
    \item $D_w \leftarrow \WMGen(D_o,\secret)$
    \item Run \A and compute $\Tilde{D_w} \leftarrow \A(D_w)$
    \item \A wins if $\Tilde{D_w} \approx D_w$ and $\WMDet(\Tilde{D_w},\secret)=reject$
\end{compactenum}
\end{definition}

\begin{definition}[Unforgeability]
An adversary only knowing the watermarked \data but not \secret is unable to convince a third party that it owns the \data. 
For every PPT algorithm \A, the chance of winning the following game is negligible:
\begin{compactenum}
     \item Compute $\secret \leftarrow \WMKeyGen(1^\lambda)$
    \item $D_w \leftarrow \WMGen(D_o,\secret)$
    \item Run \A and compute $(\Tilde{D_w},\secret') \leftarrow \A(D_w)$
    \item \A wins if $\WMDet(\Tilde{D_w},\secret')=accept$
\end{compactenum}
\end{definition}

\begin{definition}[Non-trivial Ownership]
An adversary cannot create a valid \secret for a watermarked data $D_w$, before getting any knowledge about $D_w$. 
For every PPT algorithm \A, the chance of winning the following game is negligible:
\begin{compactenum}
    \item $\secret' \leftarrow \A()$
     \item Compute $\secret \leftarrow \WMKeyGen(1^\lambda)$
    \item $D_w \leftarrow \WMGen(D_o,\secret)$
    \item \A wins if $\WMDet(D_w,\secret')=accept$
\end{compactenum}
\end{definition}

\subsection{Perceptual Hashing}\label{lsh}
Perceptual hashing algorithms generate a fingerprint for assets so that similar-looking content will be mapped to the same or similar hash values. 
More formally, we use \PHash to denote a perceptual hashing function.
To make it more comprehensible, we abuse the notation and describe \PHash as a two-input function. 
Given an input asset $D_1$ and $D_2$, the function produces a binary string as the hash value: $h=\PHash(D_1,D_2)$, where $h\in \{0,1\}^\lambda$. 
Now, assume that $D_2$ is a slightly modified version of $D_1$ (the two assets are similar) and $D_3$ denotes an asset that is different from $D_1$. 
Let $h_1$, $h_2$, and $h_3$ generated from a perceptual hash familiy $H_p$ be the hash values of $D_1$, $D_2$, and $D_3$, respectively. 
\PHash basically generates a hash value on $D_1$ while meeting the following requirements \cite{lshfarid2021overview}: (1) unpredictability of hash values: 
$Pr[H_p(D_1)=h]\approx \frac{1}{2^\lambda}, \ \forall h \in \{0,1\}^\lambda$;
(2) independence of inputs: $Pr[H_p(D_1)=h_1|H_p(D_3)=h_3]=Pr[H_p(D_1)=h_1], \ \forall h_1,h_3 \in \{0,1\}^\lambda$ ; (3) producing the similar hash values for perceptually similar assets:  $Pr[H_p(D_1)=H_p(D_2)]\approx 1 $; and (4) producing distinct values for different assets: $Pr[H_p(D_1)=H_p(D_3)]\approx 0 $.

\subsection{Digital Signature}
A digital signature is an authentication mechanism, consisting of three PPT algorithms (\SignKeyGen, \Sign, \SignVerify):
\begin{itemize}
    \item $\SignKeyGen(1^\lambda)$ generates a secret signing key $ssk$ and a corresponding public verification key $pvk$;
    \item $\Sign(ssk,m)$ produces a signature $\sigma$ of a given message $m$, using $sk$; and
    \item $\SignVerify(svk,\sigma, m)$ either \textit{accepts} (e.g., outputs $1$) or \textit{rejects} (e.g., outputs $0$) the claim that given ($m,\sigma$), $\sigma$ is a valid signature of $m$, using $svk$.
\end{itemize}

We assume a digital signature secure against adaptive existential forgery attacks. 
No PPT adversary holding $svk$ can generate a valid signature on a new message~\cite{katz2014introduction,menezes2018handbook}. 

\subsection{SGX Functionality}\label{sgxfuncdef}
We adopt the SGX functionality defined previously by Pass et al.~\cite{teeabstactPassST17} (as adopted by many TEE-based applications, e.g. \cite{zhang2016town}). 
Specifically, we use a global functionality $\FSgx$ to denote (an instance of ) an SGX functionality parameterized by a (group) signature scheme $\Sign$ as presented below. 
$\texttt{prog}_\texttt{enclave}$ denotes the SGX enclave program. 
Upon initialization, $\FSgx$ runs  $\texttt{out}:=\texttt{prog}_\texttt{enclave}.\texttt{\textbf{Init()}}$ and attests to the code of $\texttt{prog}_\texttt{enclave}$ as well as $\texttt{out}$. 
Upon a resume call with $(id, params)$, $\FSgx$ runs and outputs the result of $\texttt{prog}_\texttt{enclave}$$.\texttt{\textbf{Resume}}(\texttt{id,params})$. 

\begin{functionality}[label=sgxfunc]{$\FSgx$}
\textbf{\texttt{Embedded:}} $ssk$\\
$\texttt{prog}_\texttt{enclave}$: denotes SGX enclave program\\
\texttt{\textbf{Initialize}:}
\begin{itemize}
    \item $\texttt{out}:=\texttt{prog}_\texttt{enclave}.\texttt{\textbf{Init()}}$
    \item $\sigma_{att}:=\Sign(ssk,\texttt{prog}_\texttt{enclave}||\texttt{out})$
\end{itemize}
\textbf{\texttt{Output:}} $\langle \sigma_{att},\texttt{out}\rangle$\\
\texttt{\textbf{Resume}:}\\
On receive (\texttt{resume, id, params})\\
$\texttt{out}:=\texttt{prog}_\texttt{enclave}.\texttt{\textbf{Resume}}(\texttt{id,params})$
\end{functionality} 

\subsection{Symmetric Encryption Scheme (SE)} 
SE is a form of encryption that uses the same secret key for both encryption and decryption, which consists of three PPT algorithms, $(\Gen,\Enc,\Dec)$: 
\begin{itemize}
    \item $\Gen(1^\lambda)$ generates a secret key $k$;
    \item $\Enc(k,m)$ encrypts a message $m$ using the secret key $k$ and outputs a ciphertext $c$; and
    \item $\Dec(k,c)$ decrypts a ciphertext $c$ using the secret key $k$ and outputs a message $m$ such that $c \leftarrow \Enc(k,m)$.
\end{itemize}    
We assume a semantically secure SE scheme, i.e., the ciphertexts of two different messages are indistinguishable.

\subsection{Oblivious Transfer}\label{otdef}
Oblivious transfer (OT)  is a fundamental cryptographic primitive commonly used as a building block in MPC. 
In an OT~\cite{yadav2021survey}, a sender $S$ has two inputs $x_0$ and $x_1$. 
A receiver $R$ has a selection bit $b \in \{0,1\}$ and wants to obtain $x_b$ without learning anything else (i.e., $x_{1-b}$) or revealing $b$ to $S$. 

\subsection{Garbled Circuits (GC)}\label{garbledef}
A \emph{circuit garbling} scheme was introduced by Yao~\cite{yao86} in passive security. 
For our work, we use the abstraction presented by Bellare et al.~\cite{BellareHR12}. 
A garbling scheme consists of five algorithms: \Garble, \Encode, \GarbledEval, \Decode, and \Evaluate. 
\begin{itemize}
    \item $(\garbledCircuit, \encodingInfo,\decodingInfo) \leftarrow \Garble(1^\lambda,\circuit)$: Garbling algorithm \Garble takes an original function \circuit, describing the function $\Evaluate(\circuit,.): \{0,1\}^n \leftarrow \{0,1\}^m$ and security parameter $\lambda$ and returns garbled circuit \garbledCircuit, encoding information \encodingInfo which encodes an initial input to garbled input, and decoding information \decodingInfo which maps garbled output to final output. 
    \item $\garbledInput \leftarrow \Encode(\encodingInfo,\plainInput)$: Encoding algorithm \Encode  encodes an initial input \plainInput using encoding information \encodingInfo to garbled input \garbledInput.
    \item $\garbledOutput \leftarrow \GarbledEval(\garbledCircuit,\garbledInput)$: Garbled evaluation algorithm \GarbledEval taking the garbled circuit \garbledCircuit and garbled input \garbledInput as an input maps a garbled input \garbledInput to garbled output \garbledOutput.
    \item $\plainOutput \leftarrow \Decode(\decodingInfo,\garbledOutput)$: Decoding algorithm \Decode maps garbled output \garbledOutput to final output \plainOutput
    using decoding information \decodingInfo.
    \item $\plainOutput \leftarrow \Evaluate(\circuit,\plainInput)$: Evaluation algorithm \Evaluate, takes an initial input \plainInput and the function \circuit and returns the final output \plainOutput. 
\end{itemize}
We briefly describe here the key properties satisfied by garbling schemes. 
First, it must be correct: Given that $\circuit \in \{0,1\}^*, \lambda \in \mathbb{N}, \plainInput \in \{0,1\}^{f.n}$ and $(\garbledCircuit,\encodingInfo,\decodingInfo) \in [\Garble(1^\lambda,\circuit)]$ then $\Decode(\decodingInfo,\GarbledEval(\garbledCircuit,\Encode(\encodingInfo,\plainInput))=\Evaluate(\circuit,\plainInput)$. 
Second, it must be private: given \garbledCircuit and \garbledInput, the evaluator should not learn anything about \plainInput or \circuit except the size of \circuit and the output \plainOutput. 

\section{Formal Definition and Security Proof}\label{teeproof}
\subsection{Formal Definition}\label{formaldef}
To denote a two-party protocol between two parties, $P_1$ and $P_2$, we use the notation: 
${\footnotesize P_1[\texttt{output}_{P_1}],P_2[\texttt{output}_{P_2}] \leftarrow \texttt{Protocol(}P_1[\texttt{input}_{P_1}],}$ 
\\${\footnotesize P_2[\texttt{input}_{P_2}])}$.\\
\emph{Ownership generation} phase is formally defined as follows:

\begin{puppyprotocolnonum}[label=ownershipGen]{Ownership Generation Phase}
On input of the security parameter $\lambda$ and $D_o$ from \Ow, generate $D_w$, using \secret, and 
identifier $\idtx$ for $D_o$ and distribute $(\hshare,\pshare)$ as follows:
$$\Ow[\idtx,D_w,\secret],\HD[\idtx,D_w,\hshare],\Prov[\idtx,\pshare] \leftarrow $$
  $$\texttt{\textbf{Generate}}(\Ow[D_o,1^\lambda]),$$
\end{puppyprotocolnonum}

\emph{Verification} phase is formally defined as follows:
\begin{puppyprotocolnonum}[label=verificationFunc]{Verification Phase}
On input of $\idtx$, $D_w$, \tokenH $\hshare$ from \HD, and a database of \tokenPs \db from \Prov, 
determine validity of $D_w$ as follows:  
  $$\HD[res] \leftarrow \texttt{\textbf{Verify}}(\HD[\idtx,\hshare,D_w],\Prov[\db]),$$ 
where $res$ is `1' for valid or `0' for invalid result.
\end{puppyprotocolnonum}

\subsection{Security Definition}\label{securitydef}
We define the security of \puppy (based on Figure \ref{fig:puppy}) via the \textit{ideal and real world} paradigm~\cite{canetti}. 
Intuitively, any adversary in the real-world protocol \Ppuppy should not learn anything more than it learns from the ideal world while communicating with the ideal functionality \Fpuppy. 
\begin{itemize}
\item \textbf{Ideal World:} consists of owner \Ow, holder \HD, prover \Prov, and ideal functionality \Fpuppy. 
\item \textbf{Real World:} consists of \Ow, \HD, and \Prov. 
They run the real-world protocol \Ppuppy := (\texttt{Generate, Verify}).
\end{itemize}

\begin{functionality}[label=puppyfunc]{\Fpuppy}
\textbf{Ownership Generation:}

\begin{enumerate}
	\item receives $D_o$ from the owner \Ow.
    \item generates \secret and the tokens $\langle \hshare,\pshare \rangle$, and computes $D_w$, and $\idtx$.
	\item sends $\langle \idtx,D_w,\secret \rangle $ to \Ow, $\langle \idtx, D_w, \hshare \rangle$ to $\HD$ and $\langle \idtx, \pshare \rangle$ to $\Prov$.
	\item stores all data in the database $\db$.
\end{enumerate}
\textbf{Verification:}
\begin{enumerate}
    \item  receives $\langle \idtx, \hshare, D_w \rangle$ from \HD and \db of $(\idtx,\pshare)$ from \Prov.
	\item aborts if the following occurs and sends $\langle \abort \rangle$ to \HD:
	\begin{itemize}
	    \item $\idtx \neq \IDGen(D_w)$; or
	    \item $\idtx \notin \db$.
	\end{itemize}
	\item performs the watermark detection algorithm and sends $res$ to \HD.
\end{enumerate}
\end{functionality}

\begin{definition}[Publicly Verifiable Watermarking Protocol] \label{def:puppy}
Let \Ppuppy be a PPT protocol for publicly verifiable watermarking. We say that \Ppuppy is secure if for every non-uniform PPT real-world adversary $\Adv$, there exists a non-uniform PPT ideal world simulator \Dist such that all transcripts among parties and outputs between the real and ideal worlds are computationally indistinguishable; 
{ \small 
$$	\{\Ideal_{\Fpuppy}^{\Dist(aux)}(\secret,D_o,D_w,\lambda)\}  \nonumber \approx_c  \nonumber
	\{ \Real_{\Ppuppy}^{\Adv(aux)}(\secret,D_o,D_w,\lambda)\} \nonumber $$
	}
	\label{def:pubwm}
\noindent where $aux \in\{0,1\}^*$ is the auxiliary input and $\lambda$ is the security parameter. 
\end{definition}
\subsection{Illustation of \puppy-\Enclave}
\begin{figure}[htp]
\hspace*{-1cm}\includegraphics[scale=0.46]{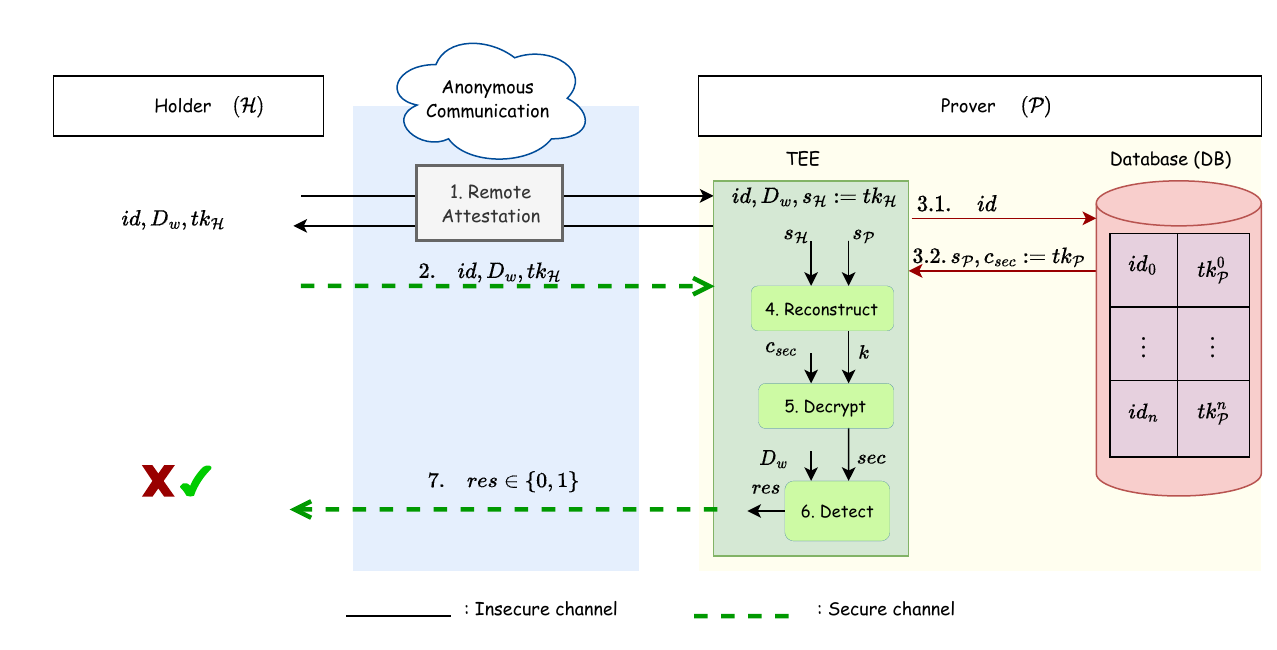}
\caption{\puppy-\Enclave Verification phase. 
}\label{fig:puppyteeVerify}
\end{figure}
\subsection{Security Proof of \puppy-\Enclave}
In this section we prove \puppy-\Enclave as define in Appendix \ref{comefficient}. 
Note that the security of \puppy-\Enclave in Section 4.6 can be proven in a very similar manner.
We assume \Ow is honest and the static malicious adversary \Adv can corrupt either \HD or \Prov, not both at the same time. 
Thus, to show the security of \puppy-\Enclave according to the Definition~\ref{def:puppy}, we need to construct ideal-world simulators for the views of \HD and \Prov. 
We assume a TEE (i.e., SGX) abstraction as a trusted third party defined by a global functionality similar to \cite{zhang2016town,tramer2017sealed}. 
Theorem \ref{thm:proofmalH} shows the security of \puppy-\Enclave with \Adv corrupting \HD and Theorem \ref{thm:proofmalP} shows one with \Adv corrupting \Prov.

\begin{theorem}\label{thm:proofmalH}
Our \puppy-\Enclave is secure according to Definition~\ref{def:puppy} against any non-uniform PPT adversary \Adv corrupting \HD assuming the underlying secret sharing scheme, and watermarking scheme are secure, the TEE on \Prov is secure (implementing secure memory, remote attestation (RA), and controlled invocation), and the cryptographic hash function is collision-resistant.

\end{theorem}
\begin{proof}
The simulator \Dist simulates honest parties in the real world, i.e., \Ow, \Prov, and the \Enclave, while \Dist simulates corrupted parties in the ideal world, i.e., \HD. 
\Dist behaves as follows: \\

\textbf{Ownership Generation Phase}:
\begin{enumerate}
    \item Once receiving a message $\langle \idtx,D_w,\hshare \rangle$ from \Fpuppy, \Dist:
    \begin{enumerate}
        \item generates a watermarked \data $D'_w$ by running the insertion algorithm based on a different \data $D'_o$ from the same space of the \data (of \Ow) and a different watermarking secret $sec' \leftarrow \WMKeyGen(1^\lambda)$ as $D'_w \leftarrow \WMGen(D'_o,sec')$;
        \item randomly chooses an element $\hsharerandom$ from the message space of $\hshare$;
        \item sends $\langle \idtx, D_w, \hsharerandom, \rangle$ to \HD in the real world; 
        \item \Dist stores all the data (i.e., for identifier, watermarked data, random share, and the original share received by \Fpuppy) in its database.
    \end{enumerate}
\end{enumerate}

\textbf{Verification Phase}:
\begin{enumerate}
    \item \Dist simulates a \Enclave in the real world, assuming it receives a valid signature key pair $\langle sk,vk \rangle$ from the manufacturer: \\
    \textbf{Remark.} Note that \Dist simulates \Enclave since \Enclave is assumed to be honest. Therefore, \Dist receives whatever \Enclave receives and outputs.
    \begin{enumerate}
        \item \Dist runs RA with \HD using $sk$ and obtains a symmetric key used for a TLS session. \\
        \textbf{Note.} Hereafter all communication with \HD (in the real world) and \Dist are assumed to be encrypted under this key (for secure channel).
    \end{enumerate} 
    \item \HD sends $\langle \tilde{\idtx}, \tilde{D_w}, \tilde{\hsharerandom} \rangle$ to \Dist over the secure channel. 
    \item \Dist checks if $\langle \tilde{\idtx}, \tilde{D_w}, \tilde{\hsharerandom} \rangle$ exists within their database (i.e., $\tilde{\idtx}=\idtx, \tilde{D_w}=D_w,$ and $\tilde{\hsharerandom} = \hsharerandom$):
    \begin{enumerate}
        \item If so, \Dist retrieves the correct share \hshare from the database and sends $\langle \idtx, D_w, \hshare \rangle$ to \Fpuppy;
        \item Otherwise, \Dist sends $\langle \idtx, D_w, \hsharerandom \rangle$ to \Fpuppy;
    \end{enumerate} 
    \item \Dist sends back the result $res$ received from \Fpuppy to \HD. 
\end{enumerate}

\begin{claim}\label{malholderclaim}
The view of \Adv (corrupting \HD) in their interaction with the \Dist is indistinguishable from the view in their interaction with the real-world honest parties.
\end{claim}

$\because$ We prove this claim via a sequence of hybrid games. 
The initial game corresponds to the real protocol, whereas the final game corresponds to the simulator \Dist described above. 
In each game, we change some steps of \puppy-\Enclave with steps that are different in the simulation above, whereas the final game corresponds to the simulator defined above without knowing the actual values of the honest parties \Prov and \Ow. 
We start our first game with the inputs of honest parties as in \puppy-\Enclave. 
Then, we send a different watermarked asset from the real watermarked asset. 
Lastly, we embed a random element chosen from the space of the real share. 
The last game corresponds to the simulator above without knowing the actual inputs of the honest parties. 
The details of the games with their reductions are below:
\begin{itemize}
  \item \textbf{Game $H_1$:}  In this game, we use the inputs of honest parties as the inputs of our simulation. Our simulation is identical to \puppy-\Enclave.
  \item \textbf{Game $H_2$:} It is same as $H_1$ except that the honest party sends a different watermarked \data $D'_w$. $D'_w$ is computed by \WMGen with inputs $D'_o$ and $\secret'$ which are randomly chosen from the \data space and watermarking key space, respectively.
  \item \textbf{Game $H_3$:} It is the same as $H_2$ except that the honest party sends a different share $\hsharerandom$ which is randomly chosen from the share space of the secret sharing scheme used in \puppy-\Enclave. 
\end{itemize}
We use the hybrid argument to show the indistinguishability of $H_1$ and $H_3$. 
If \Adv can distinguish $H_1$ and $H_3$ with a non-negligible advantage, then \Adv must distinguish the $H_i$ from $H_{i+1}$ for some $i$. 
If so, we can construct an algorithm $\B$ that breaks the underlying watermarking scheme or secret sharing scheme. 
To be specific, the distinguishability of each game has the reduction below: 
\begin{itemize}
   \item \textbf{Reduction 1:} A secure watermarking ensures that $D_w$ reveals no information about the \secret and any two watermarked data are indistinguishable without the \secret. Thus, \Adv cannot distinguish Game $H_1$ from Game $H_2$. If it can, then one can use \Adv to construct an algorithm $\B$ that breaks the security of the watermarking scheme. 
  \item \textbf{Reduction 2:} A secure secret sharing scheme ensures that the individual shares reveal no information about the secret unless they are in the access structure. If \Adv can distinguish Game $H_2$ and Game $H_3$, then it means \Adv can distinguish actual secret share from the random share so that one can use \Adv to construct an algorithm $\B$ that breaks the security of underlying secret sharing scheme. 
\end{itemize}
Since both schemes are assumed to be secure, these are the contradictions.
Therefore, there is no such $\B$ and no such \Adv, which means the execution of the real-world protocol \puppy-\Enclave is indistinguishable against the interactions with the ideal-world \Fpuppy, i.e., \puppy-\Enclave is a secure \puppy against the malicious \HD, according to the Definition~\ref{def:puppy}.
\end{proof}

\begin{theorem}\label{thm:proofmalP}
Our \puppy-\Enclave is secure according to Definition \ref{def:pubwm} against any non-uniform PPT adversary \Adv corrupting \Prov assuming the underlying secret sharing scheme and watermarking scheme are secure, the TEE on \Prov is secure (implementing secure memory, remote attestation (RA), and controlled invocation), and the cryptographic hash function is collision-resistant.
\end{theorem}
\begin{proof}
The simulator \Dist simulates the honest parties in the real world, i.e., \Ow and \HD, while it simulates the corrupted party in the ideal world, i.e., \Prov. \Dist behaves as follows:

\textbf{Ownership Generation Phase}:
\begin{enumerate}
    \item Once receiving a message $\langle \idtx,\pshare \rangle$ from \Fpuppy, \Dist:
    \begin{enumerate}
        \item randomly chooses an element $\hsharerandom$ from the message space of $\pshare$ and computes $\secret' \leftarrow \SReconst(\hsharerandom,\pshare)$; 
        \item runs the watermark insertion algorithm with $D'_o$ and $\secret'$ and obtains the watermarked \data $D_w' \leftarrow \WMGen(D'_o,\secret')$;
        \item randomly chooses an element $\psharerandom$ from the message space of $\pshare$; and
        \item sends $\langle \idtx, \psharerandom, \rangle$ to \Prov in the real world; 
        \item \Dist stores all the data in two databases: $\db$ and $\db'$.
        $\db$ consists of correct shares and the identifiers, i.e., (\pshare,\idtx), whereas $\db'$ consists of random shares and the received identifiers, i.e., ($\pshare^*$,\idtx), where $\pshare^*$ is randomly chosen from the same space as \pshare.
    \end{enumerate}
\end{enumerate}
\textbf{Verification Phase}:
\begin{enumerate}
    \item \Dist as a holder runs a RA with the \Prov in the real world to create a secure channel with \Prov.\Enclave over an anonymous communication. It aborts if the RA fails. Otherwise, it uses the secure channel to communicate with \Prov.\Enclave. 
    \item \Dist sends $\langle \idtx,D'_w,\hsharerandom \rangle$ to the \Prov.\Enclave.
    \item \Dist receives a response $res'$ from \Prov (via \Enclave on it).
\end{enumerate}
\begin{claim}\label{malproverclaim}
The view of adversary \Adv (corrupting \Prov) in their interaction with the \Dist is indistinguishable from the view in their interaction with the real-world honest parties.
\end{claim}
$\because$
We prove this claim via a sequence of hybrid games as the previous one. 
For instance, we start the first game with the inputs of honest parties as in \puppy-\Enclave. 
Then, a random share is used as input which is in the same space as the real share. 
Lastly, a different watermarked \data is used as input. 
The last game corresponds to the simulator above without knowing the actual inputs of the honest parties. 
Each game is defined as follows:
\begin{itemize}
   \item \textbf{Game $H_1$:}  In this game, we use the inputs of honest parties as the inputs of our simulation. Our simulation is identical to \puppy-\Enclave.
   \item \textbf{Game $H_2$:} It is the same as $H_1$ except that the honest party sends a different share $\hsharerandom$ which is randomly chosen from the share space of the secret sharing scheme used in \puppy-\Enclave. 
   \item \textbf{Game $H_3$:} It is the same as $H_2$ except that the honest party sends a different watermarked \data which is computed with a different \data $D'_o$ (from the same \data space of the watermarking scheme) and a different watermarking secret $\secret'$ where $\secret' \leftarrow \SReconst(\hsharerandom,\pshare)$. 
\end{itemize}

We use the hybrid argument to show the indistinguishability of $H_1$ and $H_3$. 
If \Adv can distinguish $H_1$ and $H_3$ with a non-negligible advantage, then \Adv must distinguish the $H_i$ from $H_{i+1}$ for some $i$. 
If so, we can construct an algorithm $\B$ that breaks the underlying secret sharing scheme, TEE, and anonymous communication. 
To be specific, the distinguishability of each game has the reduction below: 
\begin{itemize}
   \item \textbf{Reduction 1:}  A secure secret sharing scheme ensures that a share reveals no information about the secret.
   If \Adv can distinguish Game $H_1$ and Game $H_2$, then it means \Adv can distinguish the actual share from the random share. Thus, one can use \Adv to construct an algorithm $\B$ that breaks the security of the underlying secret sharing scheme.
  \item \textbf{Reduction 2:} A secure TEE with secure RA, controlled invocation, and isolated execution ensures: 
  \begin{itemize}
    \item the confidentiality of messages exchanged between TEE and \HD via a secure channel (established through successful RA), and 
    \item the secure execution of software (e.g., the reconstruction of \secret, the watermarking detection result $res$) run inside the TEE, and
    \item no data leakage from the secure memory inside the TEE. 
  \end{itemize}
  Hence, if \Adv can distinguish Game $H_2$ and Game $H_3$, then it means \Adv can distinguish the different watermarked \data used either from the secure channel or from the \Enclave so that one can use \Adv to construct an algorithm $\B$ that breaks the TEE guarantees. 
\end{itemize}
Since both are assumed to be secure, there is no such $\B$ nor \Adv. 
The execution of the real-world protocol \puppy-\Enclave is indistinguishable from the interactions with the ideal-world simulator \Dist. Therefore, \puppy-\Enclave is a secure \puppy against the malicious \Prov, according to the Definition~\ref{def:puppy}.
\end{proof}

\section{Other Instances of \puppy}
\subsection{Modified \puppy }\label{comefficient}
\emph{Ownership Generation Phase} 
For this construction, we use Secret Sharing to generate the two tokens, where \tokenH is denoted as $\hshare$ and \tokenP as $\pshare$. 
The shares of \secret for \HD and \Prov, \shareh and \sharep, are assigned to $\hshare$ and $\pshare$, respectively. 
\Ow generates $\idtx$ for $D_w$ as well as these two tokens, and then sends $\langle \idtx, \hshare, D_w \rangle$ to \HD, and $\langle \idtx, \pshare \rangle$ to \Prov. 
\Prov stores $(\idtx,\pshare)$ in its database, \db, located outside of the \Enclave. 
Protocol~\ref{ownership} depicts this ownership generation phase mentioned above.

\begin{puppyprotocol}[label=ownership]{Ownership Generation}
On input of $D_o$ from \Ow, the ownership of \Ow,
\begin{enumerate}
  \item \Ow computes:
 \begin{itemize}
    \item  $\secret \leftarrow \WMKeyGen(1^\lambda)$
    \item  $D_w \leftarrow \WMGen(D_o,\secret)$
    \item  $\idtx \leftarrow \IDGen(D_w)^{*}$ 
    \item  $( \shareh,\sharep) \leftarrow \SGen(\secret,\{\HD,\Prov\})$
    \item $\hshare := \shareh, \pshare := \sharep$
 \end{itemize}
  \item \Ow $\rightarrow$  \HD : $\langle \idtx,\hshare,D_w \rangle$
  \item \Ow $\rightarrow$  \Prov : $\langle \idtx,\pshare \rangle$
  \item \Prov: \db.\textbf{Store}($\idtx,\pshare$)
\end{enumerate}
\end{puppyprotocol}

\emph{Verification Phase}
During the verification phase, \HD verifies the watermark of \Ow using $D_w$ and $\hshare$ with $\idtx$.
First, \HD performs a remote attestation ($\mathsf{RA}$) on \Prov.\Enclave over the anonymous communication channel (e.g., Tor). 
It aborts if the attestation fails. 
Otherwise, \HD establishes a secure channel with \Prov.\Enclave, and from this point on, all messages between \HD and \Prov.\Enclave are sent through this secure channel. 
\HD sends $\langle \idtx$, $D_w$, $\hshare \rangle$ to \Prov.\Enclave.
Once \Prov.\Enclave receives the request, it brings \pshare from the \db using $\idtx$. 
It then reconstructs $\secret$ from the shares (via \SReconst), where the shares, \shareh and \sharep, are retrieved from the tokens, \hshare and \pshare.  
After \secret reconstruction, it runs \WMDet with $\secret$ and $D_w$, and obtains the result $res$. 
\Prov.\Enclave replies with $\langle res \rangle$ to \HD, and then erases all data including the watermarked \data, secret shares, and watermarking secret(s). 
Finally, \HD returns \texttt{Valid} if $res$ is `1' or \texttt{Invalid} otherwise.  
Protocol~\ref{verifyTEE} shows the aforementioned verification phase. 
\begin{puppyprotocol}[label=verifyTEE]{\puppy-\Enclave Verification}
On input of $\idtx$, $D_w$, and $\hshare$ from \HD, and \db from \Prov, the verification phase is as follows: 
\begin{enumerate}
  \item \HD $\xLeftrightarrow{\textbf{Tor}} \Prov.\Enclave$: \texttt{RA} (Section \ref{teeprel}) 
  \item \HD $\rightarrow$ $\Prov.\Enclave$ : $\langle \idtx,D_w,\hshare \rangle$. 
  \item $\Prov.\Enclave$ : Fetch $\langle \idtx,\pshare \rangle$ from \Prov.\db.
  \item $\shareh := \hshare, \sharep := \pshare$
  \item $\Prov.\Enclave$ : $\secret \leftarrow \SReconst(\shareh,\sharep)$.
  \item $\Prov.\Enclave$ : $res \leftarrow \WMDet(D_w,\secret)$.
  \item $\Prov.\Enclave \rightarrow \HD$ : $\langle res \rangle $
  \item $\Prov.\Enclave$: [Erase All]
  \item \HD: \texttt{\textbf{If}} $res \equiv 1$: \textbf{return} \texttt{Valid}\\
        \hspace*{0.5cm} \texttt{\textbf{Else}}: \quad  \textbf{return} \texttt{Invalid}.
\end{enumerate}
\end{puppyprotocol}
\subsection{ \puppy-\MPC}\label{com2pcefficient}
\puppy-\MPC computes the steps as described in Appendix \ref{comefficient} such that all the computations occurred in TEE are computed over a \MPC protocol. 

\section{Watermarking Schemes}\label{watermarkingschemes}

\subsection{FreqyWM}\label{freqywmdetect}
FreqyWM \cite{isler22} represents a category of watermarking schemes that is capable of inserting a watermark into a data set regardless of data type. 
After generating a histogram (based on the frequency of appearance of repeating tokens) from the data set, FreqyWM creates a pair for each entry and each other entry and calculates a hash of the pair using a secret key $K$, modulus of a secret integer $z$, resulting in a value $m$. 
FreqyWM then calculates the difference of the pair frequency values modulus of $m$ and runs the result through an optimization algorithm to find the best entry pair that allows the final result to be 0 by modifying the frequency value of entries. 
The list of chosen pairs, $K$, and $z$ are the secret values for FreqyWM. 
We implemented the watermark detection algorithm in \texttt{C} as FreqyWM does not have an open-source implementation. 
The total size of the combined secret values is about 2 kB. 
FreqyWM detection is presented in Algorithm \ref{WMDetect} (see \cite{isler22} for details). 
It takes four inputs: 1) a watermarked \data $D_w$; 2) a watermarking secret \secret which is a list as $\secret=\{L_{wm},K,z\}$, where $L_{wm}$ is the list of watermarked pairs ($\uL_i, \uL_j$), $K$ is a high entropy value, and $z$ is an (modulo) integer; 3) a \textit{threshold} $t$ to decide if a certain pair is watermarked; and 4) a threshold $k$ representing the \textit{minimum number of watermarked pairs} required to conclude whether $D_w$ is a watermarked dataset. 

\begin{algorithmdes}[label=WMDetect]{FreqyWM Watermark Detection}
({\color{blue} \it Computations outside of 2PC from Protocol~\ref{verifygc} represented in blue.})

\KwIn{$D_w,\secret=\{L_{wm},K,z\},k,t$}
\KwOut{$accept/reject$}
{\color{blue}$D^{hist}_w=\textbf{Preprocess}(D_w)$}\;\\
$count=0$, $result=reject$\;\\
\ForEach{$\{\uL_i,\uL_j\} \in L_{wm}$}{
 \If{\textbf{Found($\uL_i,\uL_j$, $D^{hist}_w$)}}{
    $s_{ij}= Hash( \uL_i|| \uL_j||K) \mod z$\;
	
		\If{$(f_{i} - f_{j}) \mod s_{ij} \leq t$ }{
		   $count++$\;
		}}
}
	\If{$count \geq k$ }{
		 $result=accept$\;
		}
	\KwResult{result}
\end{algorithmdes}

\subsection{OBT-WM}
OBT-WM \cite{shehab2007watermarking} represents a category of watermarking schemes for numerical databases. 
OBT-WM first partitions the database based on a secret key $K$, primary keys, and a predefined number of partitions $num\_partitions$. 
Once the partitions are chosen, the values in the database are modified to encode a predefined watermark (a list of bits) $wm$ by solving a constrained optimization problem. 
The watermark $wm$, $num\_partitions$, and $K$ are the secret values for OBT-WM. 
Similar to FreqyWM, OBT-WM also does not have an open-source code, so the watermark detection algorithm was implemented in \texttt{C}. 
The total size of the combined secret values is approximately 80 bytes. 

\subsection{IMG-WM}
IMG-WM \cite{lou2007copyright} represents a category of image watermarking schemes. 
IMG-WM assumes a monotone image (i.e., pixel values are either 0 or 255). 
During the watermark insertion phase, it first uses a Discrete Wavelet Transform (DWT) technique and divides the original image into ten sub-bands. 
From the ten sub-bands, it extracts the low and middle sub-bands $L$ and $M$ and computes new coefficients of the low sub-band $L^\prime$ using a secret key $K$, $L$, and $M$. 
Finally, a random share $share1$ is calculated according to the difference between $L$ and $L^\prime$ and the watermark image, and another share $share2$ is generated by XOR-ing $share1$ with the watermark image. 
During the watermark detection phase, $share1$ and $share2$ are XOR-ed and checked for matching the watermark. 
$share1$, $share2$, and the watermark image are the secret values for IMG-WM. 
An open-source implementation of IMG-WM is provided at~\cite{imagewatermark} that uses OpenCV~\cite{opencv} for the detection phase. 
Each share is about 80 kB in size, and therefore the total size of the combined secret values is approximately 240 kB. 

\subsection{DNN-WM}
DNN-WM \cite{adi2018turning} represents a category of machine learning model watermarking schemes. 
DNN-WM uses a well-known attack called ``backdooring'' to purposefully train the model to categorize a set of predefined images (called $triggerset$) with a wrong label (called $watermark\_label$). 
$triggerset$ and $watermark\_label$ are the secret values for DNN-WM. 
An open-source implementation of DNN-WM is provided at~\cite{nnwatermark} that uses PyTorch~\cite{pytorch} for the insertion/detection phases. 
The size of the sample $triggerset$ in the implementation is about 17 MB and $watermark\_label$ is 4 kB, therefore the total size of the secret value results in around 17 MB. 

\end{document}